\documentclass{article}
\pdfoutput=1

\usepackage{arxiv}

\usepackage[utf8]{inputenc} 
\usepackage[T1]{fontenc}    

\usepackage{url}            
\usepackage{booktabs}       
\usepackage{amsfonts}       
\usepackage{nicefrac}       
\usepackage{microtype}      
\usepackage{lipsum}
\usepackage{amsmath,amssymb,amsfonts,mathtools}
\usepackage[bookmarks=false]{hyperref}
\usepackage{float}
\usepackage{comment}
\usepackage{graphicx}
\usepackage{tabularx}
\usepackage{adjustbox}
\usepackage{subfig}
\usepackage{blindtext}
\usepackage{array}
\usepackage{mdwmath}
\usepackage{mdwtab}
\usepackage[xcdraw]{xcolor}
\usepackage[normalem]{ulem}
\usepackage{stackengine}
\usepackage{booktabs}
\usepackage{varwidth}
\usepackage[belowskip=-15pt,aboveskip=0pt]{caption}
\usepackage[binary-units=true]{siunitx}
\usepackage{breakcites}
\usepackage{etoolbox}

\usepackage[subfigure,titles]{tocloft}
\addtocontents{lof}{\cftpagenumbersoff{figure}}

\renewcommand\cftfigpresnum{\bfseries Figure }

\newlength\mylength
\makeatletter
\def\doublecaption{%
   \ifx\@captype\@undefined
     \@latex@error{\noexpand\caption outside float}\@ehd
     \expandafter\@gobble
   \else
     \refstepcounter\@captype
     \expandafter\@firstofone
   \fi
   {\@dblarg{\@doublecaption\@captype}}%
}
\long\def\@doublecaption#1[#2]#3{%
  \par
  \addcontentsline{\csname ext@#1\endcsname}{#1}%
    {\protect\numberline{\csname the#1\endcsname}{\ignorespaces #2\newline#3}}%
  \begingroup
    \@parboxrestore
    \if@minipage
      \@setminipage
    \fi
    \normalsize
    \@makecaption{\csname fnum@#1\endcsname}{\ignorespaces #3}\par
  \endgroup}
\makeatother

\title{High Precision Hybrid RF and Ultrasonic Chirp-based Ranging for Low-Power IoT Nodes.}

\author{
  Bert Cox \\
  ESAT-DRAMCO Research Group\\
  KU Leuven\\
  Gebroeders De Smetstraat 1, 9000 Gent, Belgium \\
  \texttt{bert.cox@dramco.be} \\
   \And
 Liesbet Van der Perre \\
  ESAT-DRAMCO Research Group\\
  KU Leuven\\
  Gebroeders De Smetstraat 1, 9000 Gent, Belgium \\
  \texttt{liesbet.vanderperre@dramco.be} \\
  \AND
  Stijn Wielandt \\
  Earth and Environmental Sciences Area\\
  Lawrence Berkeley National Laboratory\\
  1 Cyclotron Rd, Berkeley, CA 94720, United States \\
  \texttt{stijnwielandt@lbl.gov} \\
  \And
  Geoffrey Ottoy\\
  ESAT-DRAMCO Research Group\\
  KU Leuven\\
  Gebroeders De Smetstraat 1, 9000 Gent, Belgium \\
  \texttt{geoffrey.ottoy@dramco.be} \\
  \And
  Lieven De Strycker\\
  ESAT-DRAMCO Research Group\\
  KU Leuven\\
  Gebroeders De Smetstraat 1, 9000 Gent, Belgium \\
  \texttt{lieven.destrycker@dramco.be} \\
}

\begin{document}
\maketitle

\begin{abstract}
Hybrid acoustic-RF systems offer excellent ranging accuracy, yet they typically come at a power consumption that is too high to meet the energy constraints of mobile IoT nodes. We combine pulse compression and synchronized wake-ups to achieve a ranging solution that limits the active time of the nodes to 1\,ms. Hence, an ultra low-power consumption of \SI{9.015}{\micro \watt} for a single measurement is achieved. Measurements based on a proof-of-concept hardware platform show median distance error values below 10\,cm. Both simulations and measurements demonstrate that the accuracy is reduced at low signal-to-noise ratios and when reflections occur. We introduce three methods that enhance the distance measurements at a low extra processing power cost. Hence, we validate in realistic environments that the centimeter accuracy can be obtained within the energy budget of mobile devices and IoT nodes. 
The proposed hybrid signal ranging system can be extended to perform accurate, low-power indoor positioning.
\end{abstract}


\section{Introduction}
\label{intro}
Accurate positioning of users and devices plays a major role in the growing number of location-aware applications.
In time-based localization ranging systems, acoustic signals are inherently interesting candidates for precise ranging thanks to their relatively low propagation speed. Unlike RF-based systems, they do not require high processing speeds, nor the same level of synchronization accuracy. However, they are receptive to environmental and room characteristics, such as temperature, relative air velocity, reflection and diffraction, impacting the accuracy of the measurements.~\cite{Kuttruff2017}.
Hybrid signal ranging combines the advantages of both wave types: an RF signal is used as a reference and the time difference with the slower propagating (ultra)sound signal is used to calculate the distance. In hybrid RF/acoustic positioning, two classes of techniques have been proposed: indirect positioning and self positioning. An example of the former is the Active Bat Local Positioning System~\cite{ActiveBat}. The base stations are attached to the ceiling and periodically broadcast a radio message containing a single identifier, causing the corresponding mobile node to emit a short encoded ultrasound pulse. The position is then calculated at a central point using the Time of Arrival (ToA) at the different beacons. The second technique is used by the Cricket System~\cite{Cricket}. Here, the base stations simultaneously emit ultrasonic and radio pulses for Time Difference of Arrival (TDoA) calculations. The mobile device calculates its own position, which ensures its privacy. Several hybrid signaling studies have focused on obtaining high accuracy with as little infrastructure as possible~\cite{Medina, Khyam}, paying little or no attention to the energy constraints of mobile devices in location-aware system designs. To address the latter, short-range, backscattering-based approaches have been proposed with high precision but complex system architecture.~\cite{Zhao, Cox2}\newline
We present three main contributions in this paper. The first contribution is a novel hybrid RF-acoustic signaling that performs a just-in-time wake up of the nodes. This concept enables mobile receiver entities with ultra low power consumption compared to conventional schemes. Secondly we propose fast and lightweight algorithmic solutions to resolve incorrect measurements due to reflective acoustic signals. The third contribution is the realization of an experimental set-up that was used to validate the results in real-life proof of concept. \newline
This paper is further organized as follows. The next section introduces the chirp-based ranging concept, focusing on high accuracy and ultra-low power consumption. In Section~\ref{sec:experiments}, the proposed system's accuracy is assessed based on three indoor environments with different room characteristics. The next section compares three low complexity solutions to enhance the accuracy.
We present our experimental setup, hardware design and measurement results in Section~\ref{sec:results}. The last section of this paper summarizes the conclusions and discusses the potential future work.

\section{Methods}
\subsection{Ultra Low-Power Hybrid Acoustic-RF Ranging Concept}
A straightforward approach for hybrid RF/acoustic -- also used in the Cricket system~\cite{Cricket} -- ranging involves the simultaneous transmission of the signals. Hence at the receiver side, the RF signal arrives quasi-instantaneous and serves as a time reference. Consequently, the propagation delay of the acoustic signal is measured to calculate the distance between transmitter and receiver. While this method offers high precision, it is not well suited for low-power devices as both sides of the link require significant energy: 
\begin{itemize}
    \item The transmitter needs to power a loudspeaker, which is approximately 150 times more power hungry than transmitting RF \cite{Li_Xuesheng} \cite{BLE_Power}. Hence the Active Bat topology ~\cite{ActiveBat}, where the nodes emit ultrasonic signals, is not appropriate for energy constrained nodes.
    \item The receiver needs to stay powered on for a relatively long time waiting for the acoustic signal to arrive, e.g. 30\,ms for an operation area with a radius of about 10\,m.
\end{itemize}
 We propose an alternative solution that addresses the above challenges by:
 \begin{enumerate}
    \item First starting the transmission of an acoustic signal only. This signal has a predetermined duration and it is modulated in order to enable the extraction of delay information later on. The prototype presented in this paper uses a chirp signal.
    \item Transmitting the RF signal at the end of the transmission of the acoustic signal, to wake up all receivers simultaneously, for a short duration only. 
\end{enumerate}
 Note that acoustic wake-up sensors~\cite{VM1010, AcousticWakeup} can also be considered to avoid a long 'on-time' of nodes waiting for the signal to arrive. However, these typically generate frequent false wake-ups caused by ambient sounds, decreasing the energy efficiency and system accuracy drastically.
Another convenient strategy could be to broadcast audio at fixed time intervals. In these scenarios, clock drift should be countered by performing timing synchronization between the transmitting and receiving side, leading to a more complicated and power hungry system architecture. Our proposed system can transmit the hybrid signals at its own convenience and is clock drift independent due to the RF-signals both acting as a time reference and communication backbone. Advantages of this system are the reduction of the awake times to milliseconds, the prevention of false wake-ups and the opportunity to use ultrasonic sound signals, enabling human unaware, acoustic positioning and easy synchronization with RF.

\begin{figure}[!htb]
  \centering
    \includegraphics[width=0.75\textwidth]{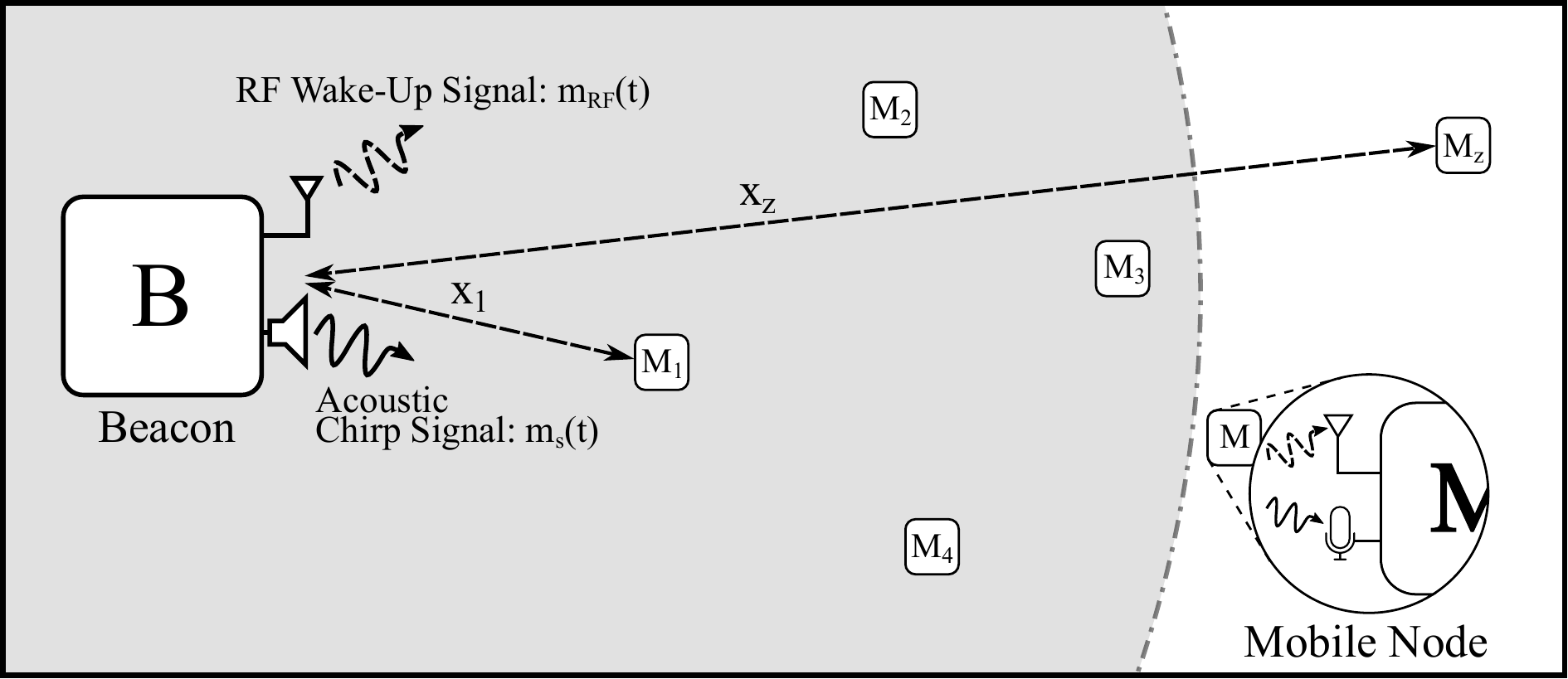}
     \vspace*{10pt} 
  \doublecaption[Single beacon ranging system setup.]{Ranging system setup. The beacon periodically transmits a sound signal. All ultra low-power nodes are woken up and synchronized on basis of the RF signal. }
  \label{Fig:SystemSetup}
\end{figure}

\subsection{Hybrid RF-ultrasonic System} \label{ss:systemOverview}

The proposed hybrid RF/ultrasonic ranging system concept is shown in Fig.~\ref{Fig:SystemSetup}. In its generic form, it consists of a single beacon (B) and one or more mobile nodes ($M_x$). The beacon is able to wake up all the mobile nodes simultaneously by a single RF signal ($m_{RF}(t)$). A distance measurement is performed as follows: 
\begin{enumerate}
    \item The beacon starts broadcasting an audio signal ($m_{s}(t)$) with a certain duration ($\tau_{tx}$) at starting time ($T_0$).
    \item At a given time ($T_A \geq T_0$), all mobile nodes wake up simultaneously for a short time ($\tau_{rx}$) and receive, depending on their distance to the beacon, a specific part of the delayed, distorted audio broadcast.
    \item Pulse compression is performed on the smaller, received audio snippet resulting in a distance estimation.
\end{enumerate}

\begin{figure}[!htb]
  \centering
    \includegraphics[width=0.65\textwidth]{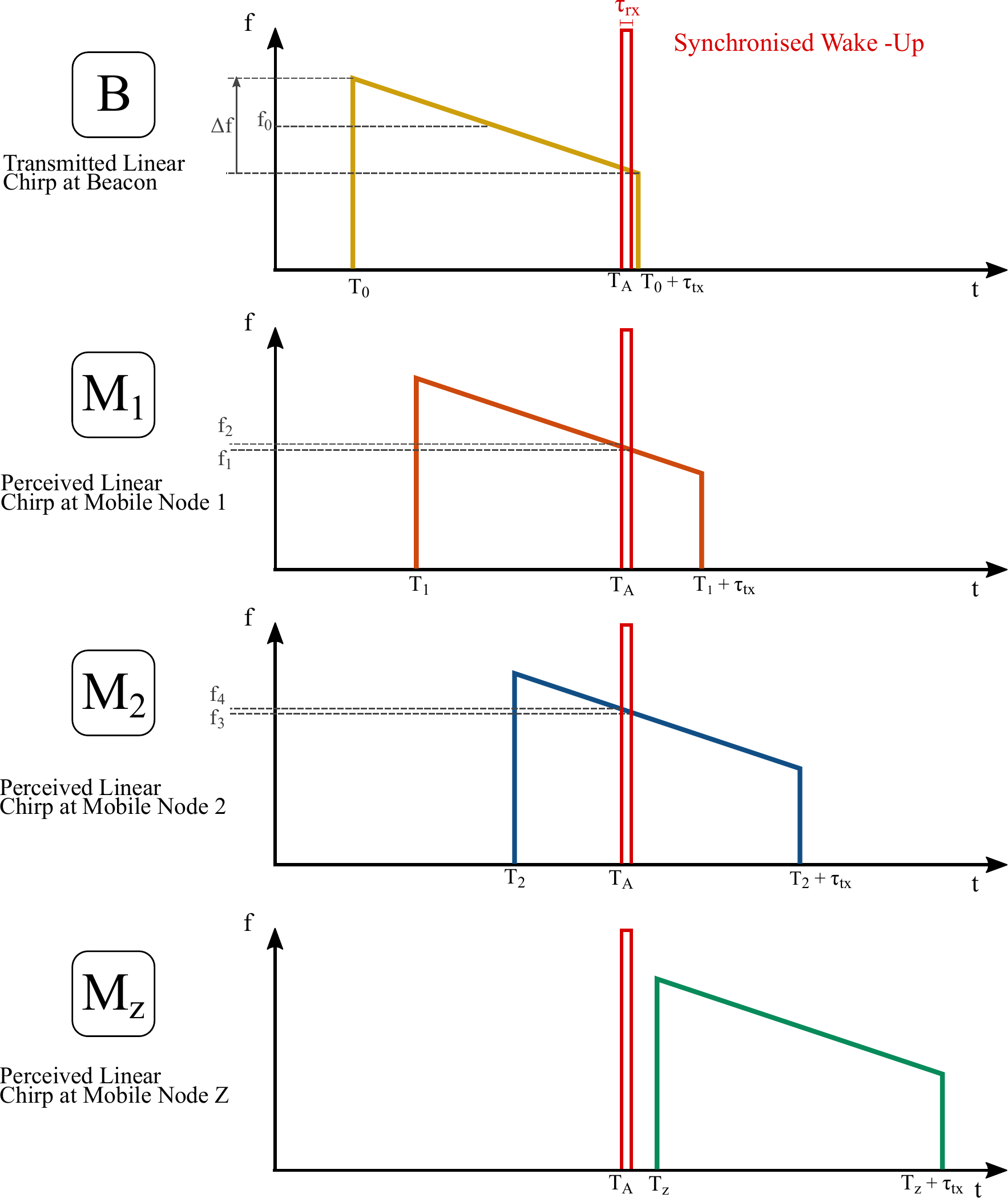}
    \vspace*{10pt} 
  \doublecaption[Timing overview of the transmitter and three mobile nodes.]{Timing overview of the transmitter and three mobile nodes. Mobile node 1 and 2 are within the range of the beacon, Mobile node 3 is not. The distances to the transmitter are calculated based on the received sound chirp.}
  \label{Fig:TimingNodes}
\end{figure}

Fig.~\ref{Fig:TimingNodes} shows a timing diagram for an exemplary case with three mobile nodes. This timing overview illustrates the difference between the proposed concept and conventional hybrid RF/acoustic TDOA systems. As all mobile nodes wake up at the same time, the ranging information is comprised in the received audio signals ($\Delta f_1$ for M1, $\Delta f_2$ for M2) at the wake-up time of the mobile nodes ($T_A$ in all cases). The restricted awake time reduces the power consumption, it impacts, however, the accuracy of the measurements. This accuracy is limited by the duration of the reception window $\tau_{rx}$ and the perceived frequency swing $\Delta f$.

\begin{figure}[!htb]
  \centering
     \vspace*{10pt} 
    \includegraphics[width=0.75\textwidth]{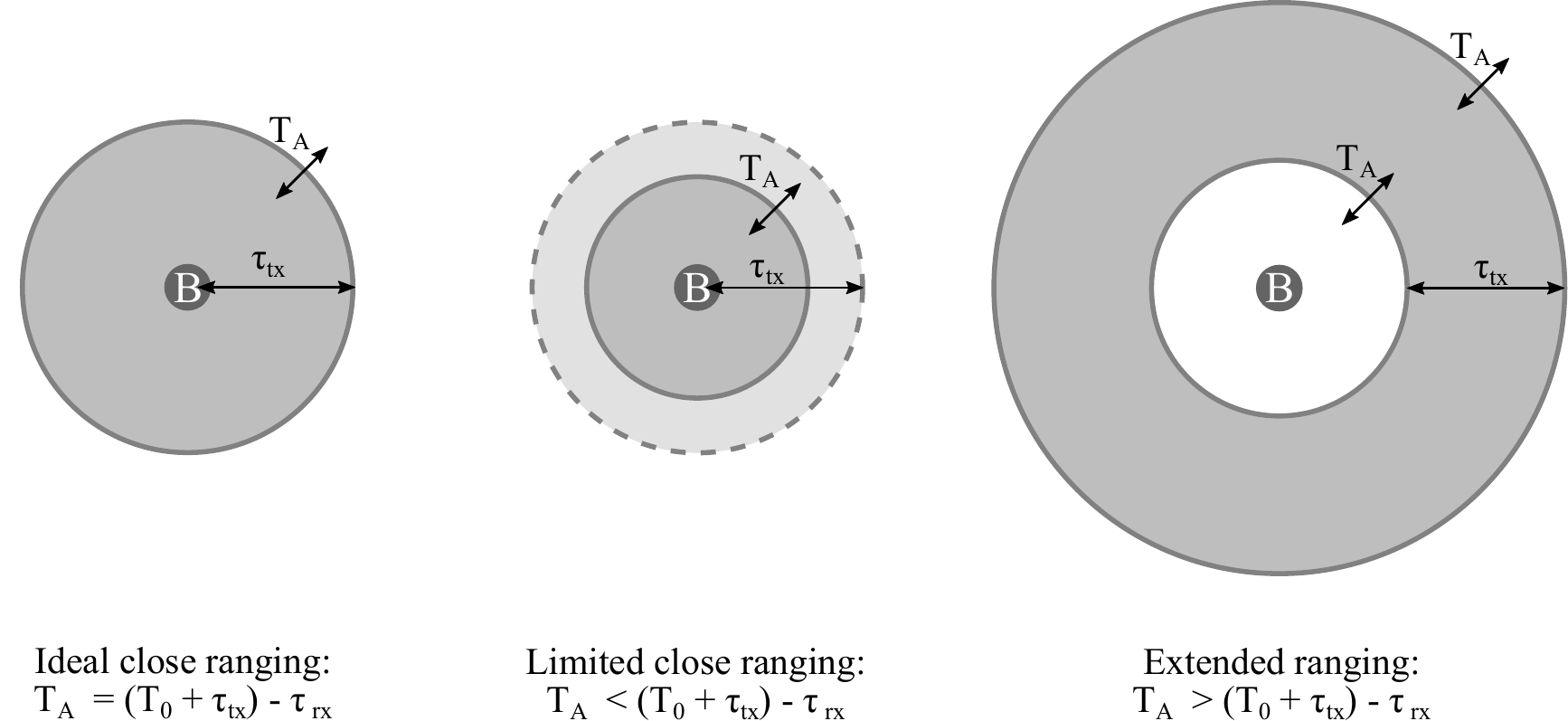}
    \vspace*{10pt} 
  \doublecaption[Three ranging scenarios.]{Three ranging scenarios at different RF wake-up signal times. 
  }
  \label{Fig:RangingWakeUp}
\end{figure}

The ranging coverage of the system\footnote{This is an area around a beacon, in which mobile nodes can measure their distance to that beacon.} is determined by the audio broadcast duration at the transmit side ($\tau_{tx}$) and the speed of sound. For example, for a sound signal with a duration of 30\,ms and a speed of sound of $v_s$ = 340\,m/s, the coverage is limited to 10.2\,m. Fig.~\ref{Fig:RangingWakeUp} illustrates the three possible ranging scenarios.

The first scenario depicts the standard operation. Here, the receivers should wake up as late as possible. More specific, the RF wake-up signals is sent at the very end of the acoustic signal, i.e., $T_A = (T_0 + \tau_{tx}) - \tau_{rx}$. $M_z$ (see Fig.~\ref{Fig:SystemSetup} and Fig.~\ref{Fig:TimingNodes}), is too far away from the beacon to receive the sound signal when it wakes up, since  $( \frac{x_z}{v_s} > T_A-T_0)$. It is therefore incapable of calculating its distance to the beacon.

The second scenario illustrates what happens when the RF wake-up signal is sent earlier during the audio broadcast. This reduces the maximum ranging coverage.

A last scenario shows what happens if the RF-awake signal is sent after the audio broadcast. Here the receivers close to the beacon are incapable of calculating their distance as they do not receive a direct audio signal during their wake-up period. On the contrary, the maximum distance to the beacon is increased.

The sampled audio in the awake state can be used for local processing (self positioning) or can be transmitted to a central unit (indirect positioning). Two-dimensional positioning of the mobile node can be achieved by adding at least two more beacons to the system in Fig. \ref{Fig:SystemSetup}.  The acquired, relative distances to these beacons can be used in multilateration or other geometric models to find the position of the mobile nodes~\cite{Munoz2009}. Identification of the sound signal is crucial here, and multiple access techniques (FDMA, TDMA, CDMA, etc. ) as proposed by~\cite{MaTDMA, FDMACiochina, CDMA_Pompili} can provide an approach to achieve this. These schemes, however, are outside the scope of this paper which focuses on the proposed ranging system in reverberant and noisy environments. 

\subsection{Pulse Compression}
Autocorrelation is used to perform fast distance calculations using small data sets, as described in~\cite{Parrilla, Marioli}, and exploited in~\cite{Nakahira, Hammoud}. A linear chirp is used as audio broadcast signal for two reasons. First of all, mobile nodes with different distances to the beacon, will measure other parts of the chirp signal. They can rely solely on the measured frequency shift $\Delta f$ to compute their distance to the beacon. Secondly, when processing, chirps remain very well correlated over Doppler shifts~\cite{Cook} and have compressed inter-correlation signals~\cite{Lazik}.

\begin{figure}[!ht]
   \centering
   \subfloat[][]{\includegraphics[width=.48\textwidth]{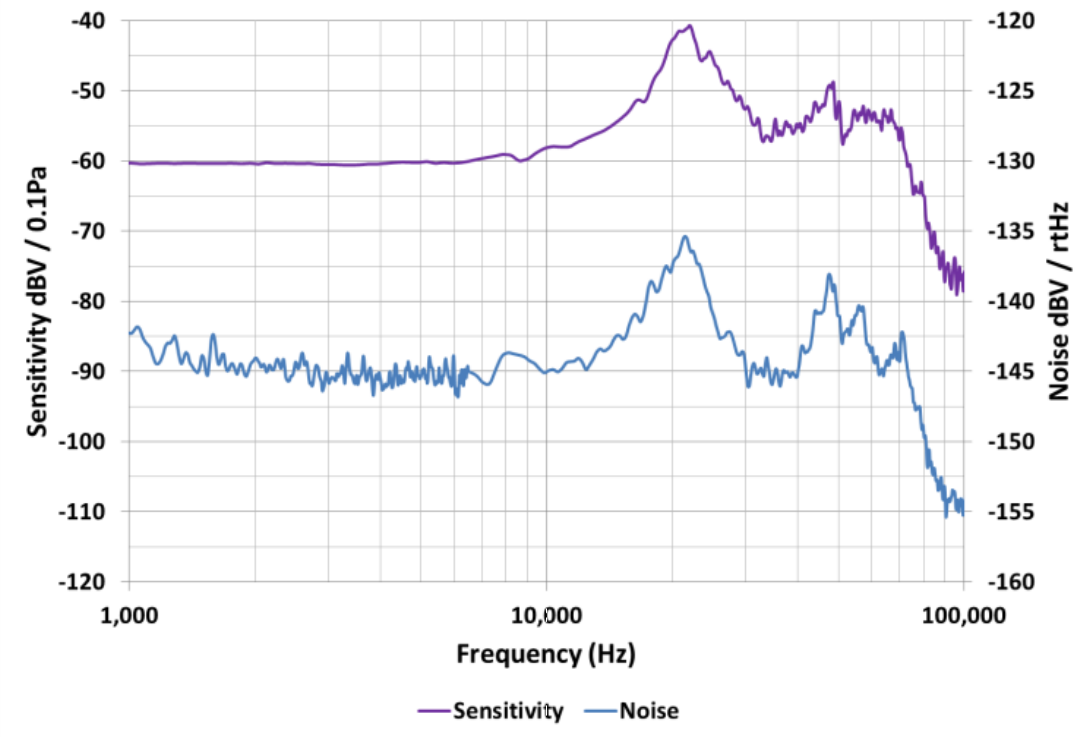}\label{Fig:SPUMEMS} }\quad
   \subfloat[][]{\includegraphics[width=.48\textwidth]{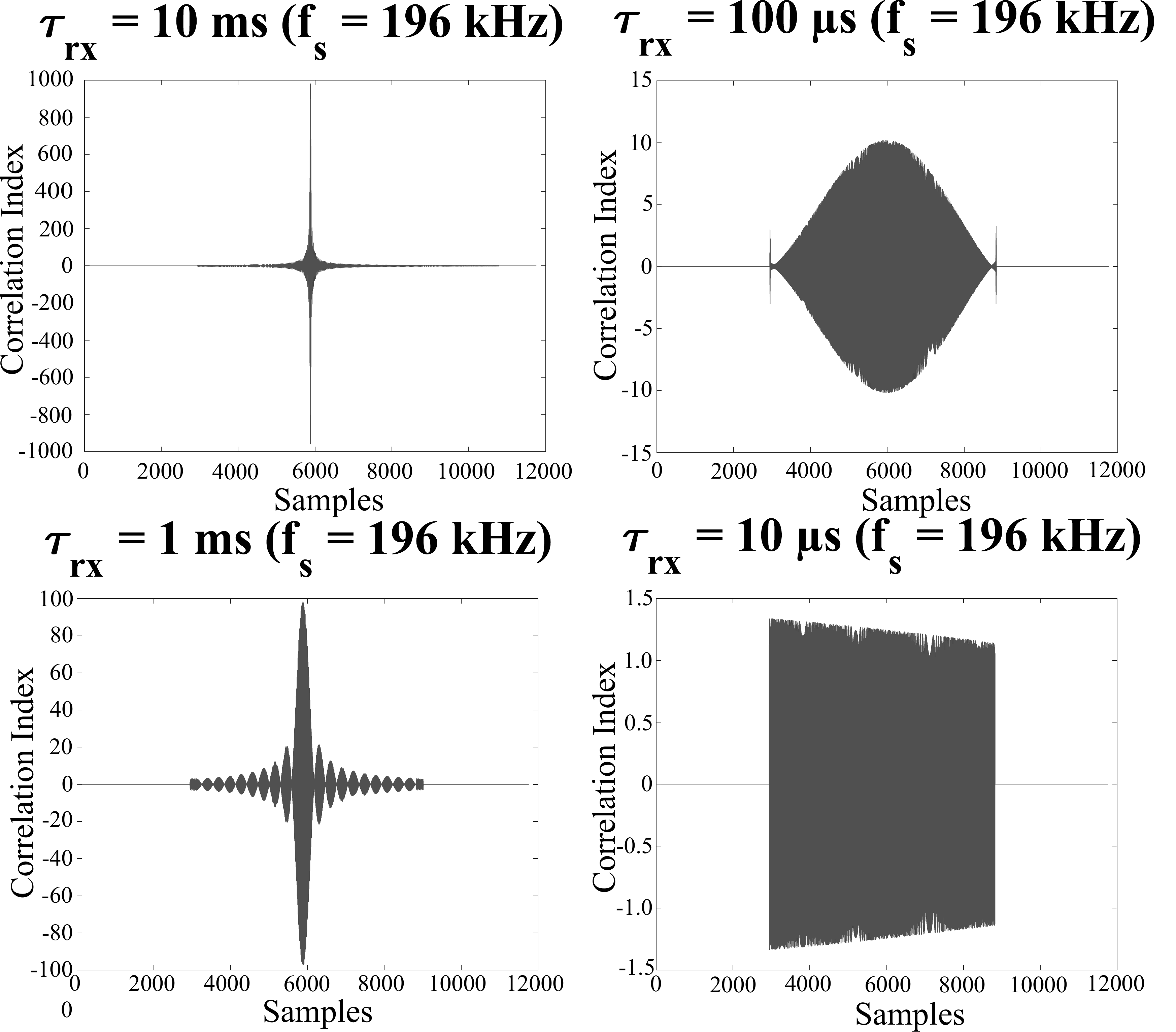}\label{Fig:PulseCompression}}\\
   \vspace*{10pt} 
   \doublecaption[MEMS microphone frequency response and pulse compression at different wake times.]{(a) SPU1410LR5H typical high frequency response and noise. \cite{SPU1410} (b) pulse compression of chirp with $\Delta f_{tx} =$ 20\,kHz and constant pulse duration $\tau_{tx} = 30$\,ms at different awake time durations ($\tau_{rx}$).}
    \vspace*{10pt} 
\end{figure}

In linear chirps, using cross correlation/autocorrelation is a form of pulse compression. A linear chirp, $s_c(t)$, can be described as:
\begin{equation}
s_c(t)=
\begin{dcases}
A\,e^{i2\pi \, \left( \left(f_0 - \frac{\Delta f}{2} \right)t + \frac{\Delta f}{2 \tau} \, t^2 \right)} , & \text{if } 0 \leqslant t < \tau_{rx} \\
 0, & \text{otherwise} \, ,
\end{dcases}
\end{equation}

where $\tau_{rx}$ is the pulse duration, $A$ the amplitude of a rectangle window function, $f_0$ is the carrier frequency and $\Delta f$ is the nominal chirp bandwidth. 
The equation for the instantaneous frequency $f(t)$ shows this linear ramp of the chirp:
\begin{equation}
f(t) = \frac{1}{2\pi} \left[\frac{d\phi}{dt} \right]_t = f_0 - \frac{\Delta f}{2} + \frac{\Delta f}{\tau}t \, ,
\end{equation}
where $\phi(t)$ is the phase of the chirped signal.\newline
Cross correlation between the transmitted and received signal can be achieved by convolving the received signal with the conjugated and time-reversed transmitted signal:
\begin{equation}
\langle s_c , s_c \rangle(t) = \int_{-\infty}^{+\infty} s_c^\star (\tau)\, s_c(t+\tau) d\tau \, .
\end{equation}
It can be shown \cite{Hein} that the autocorrelation of a chirp signal is given by:
\begin{equation}
\langle s_c , s_c \rangle(t) = A^2\tau \, \Lambda \left(\frac{t}{\tau_{rx}}\right) sinc\left[\Delta f t \, \Lambda\left(\frac{t}{\tau_{rx}}\right) \right] e^{2 i \pi f_0 t} \, , 
\end{equation}
with $\Lambda$ a triangle function, with a value of $0$ on $[\text{-}\infty, \text{-}\frac{1}{2}]\cup[\frac{1}{2},\infty]$ and linearly increasing on $[\frac{-1}{2},0]$ where it has its maximum 1, and then decreasing linearly on $[0,\frac{1}{2}]$. Around the maximum, this function behaves like a cardinal sine, with a $\text{-}3$\,dB width of  $\tau' \approx \frac{1}{\Delta f_{rx}}$. For common values of $\Delta f_{rx}$, $\tau'$ is smaller than $\tau_{rx}$, hence the name pulse compression. \newline
The pulse compression ratio can be described as the ratio between the received pulse and the compressed pulse duration:
\begin{equation} \label{Eq:CompressionRatio}
\frac{\tau_{rx}}{\tau'} = \tau_{rx} \, \Delta f_{rx} = \tau_{rx}^2 \, \frac{\Delta f_{tx}}{\tau_{tx}} \, .
\end{equation}
This equation can be rewritten as the time bandwidth product and is generally larger than 1. As the energy of the signal is kept constant when pulse compression is performed, the energy gets concentrated in the main lobe of the cardinal sine, resulting in an SNR-gain proportional to the compression ratio.\newline
 There are three parameters in Equation (\ref{Eq:CompressionRatio}) that can increase the SNR-gain and inherently result in more accurate distance calculations. 
 \begin{enumerate}
     \item The nominal chirp bandwidth ($\Delta f_{tx}$), is limited by the frequency response of the ultrasonic microphone or speaker. Typically for low cost, low-power mobile acoustic nodes, MEMS microphones are used, e.g. Knowles SPU1410LR5H \cite{SPU1410}. Fig. \ref{Fig:SPUMEMS} shows the frequency and noise response of this MEMS microphones. A theoretical -3\,dB bandwidth of 75\,kHz can be deducted from this picture. In practice, this bandwidth will be lower after amplification.
     \item For an optimal compression ratio, the audio broadcast duration ($\tau_{tx}$) should be as low as possible. However, limiting this parameter would decrease the coverage range, as explained in Fig. \ref{Fig:RangingWakeUp}.
     \item The last parameter is the receiver awake time ($\tau_{rx}$). This awake time should be set with care. On the one hand, there is a direct relationship with the time-bandwidth. Increasing $\tau_{rx}$ improves the SNR and accuracy quadratically. On the other hand, the receiver awake time should be kept as low as possible to restrain the power consumption for a single measurement.
 \end{enumerate}
  As a visual example, the pulse compression of a chirp with a nominal chirp bandwidth of $\Delta f_{tx} = $20\,kHz and a constant pulse duration $\tau_{tx} = 30$\,ms for different awake time durations are depicted Fig. \ref{Fig:PulseCompression}. This picture shows that correct maxima at a sufficient resolution can be derived from the pulse compression calculations when the awake time is limited to 1\,ms.

\section{Validation: Simulation-based Performance Analysis}\label{sec:experiments}
\subsection{Simulation Framework}

To assess the performance of the proposed ranging system under realistic noise conditions and in reflective indoor environments, we established a simulation framework. Specifically, the robustness of the inter-correlation performance in different positions, possibly suffering from reverberation, has been investigated. The simulation framework is based on the Image Source Model (ISM), which has been extensively used in room acoustics because of its fast processing speed with accurate results in case of box shaped rooms \cite{AllenBerkley}. The Allen and Berkley algorithm used in this ISM, calculates the Room Impulse Response (RIR) at the receiver's position using a time-domain image expansion method, where wall reflections are replaced by virtual sources.  All RIR-calculations were performed in a 6\,x\,4\,x\,2.5\,m room with rigid walls.  The sound source is positioned slightly off the room center at a height of 1\,m. This is done to prevent sweeping echoes \cite{DeSena} as they occur in perfect cube shaped boxes due to the orderly time-alignment of high-order reflections. Both the sound source and the receiver positioned in this room are perfectly omnidirectional. The absorption coefficients of the walls are kept uniform over all 6 planes. This is accurate as long as the wavelength of the sound is small relative to the size of the reflectors \cite{scheibler:2017}. The speed of sound is kept as a constant (340\,m/s) implying a uniform room temperature of less then 20$^\circ$\,C. The sample frequency is set to 196\,kHz which is smaller than the common values offered by off the shelf microcontroller boards (nRF52832) and larger than the Nyquist frequency of the ultrasonic MEMS microphone maximum frequency (75\,kHz). The audio broadcast duration $\tau_{tx}$ is fixed to 30\,ms, implying that every possible sensor position in the simulation environment receives a sound signal during its wake up period.\newline
We performed two types of simulations in this acoustic shoe box. The first type is a Monte Carlo simulation to test the influence of signal bandwidth and additive colored noise on the accuracy of the pulse compression technique. During these simulations, the impact of the room characteristics, such as reflection, scattering and reverberation, are kept as low as possible. It is the second type of shoe box simulations that investigates the impact of the room's characteristics, by creating simulation environments in which 600 microphones are distributed.

\subsection{Performance Assessment}
\begin{figure}[!htb]
  \centering
    \includegraphics[width=0.95\textwidth]{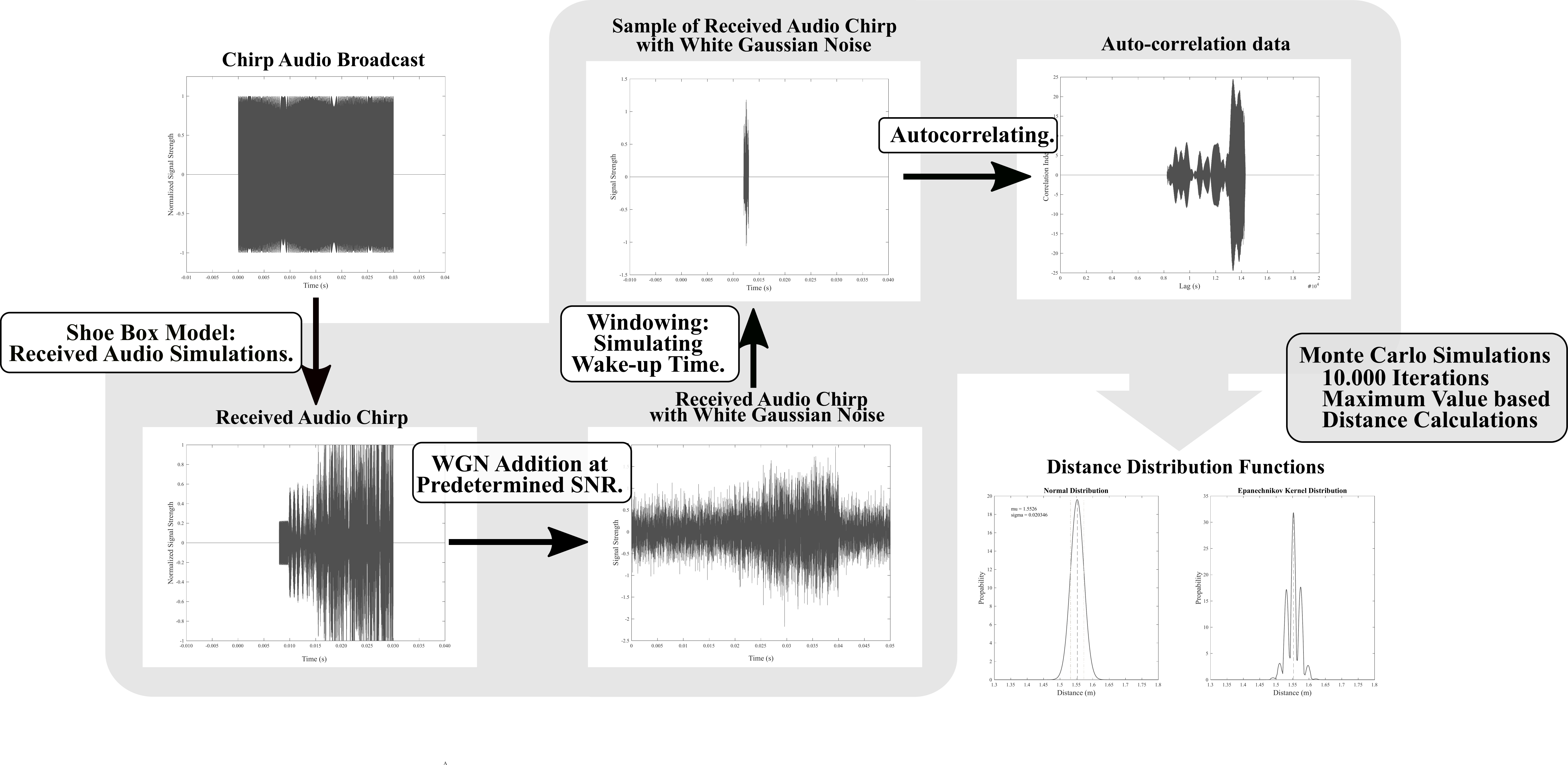}
  \doublecaption[Monte Carlo overview.]{Overview of the preprocessing, Monte Carlo simulations and postprocessing. The actual Monte Carlo simulations are repeated 10 000 times.}
  \vspace*{10pt} 
  \label{Fig:MonteCarloOverview}
\end{figure}

\begin{figure}[!htb]
\vspace*{10pt}
  \centering
    \includegraphics[width=1.0\textwidth]{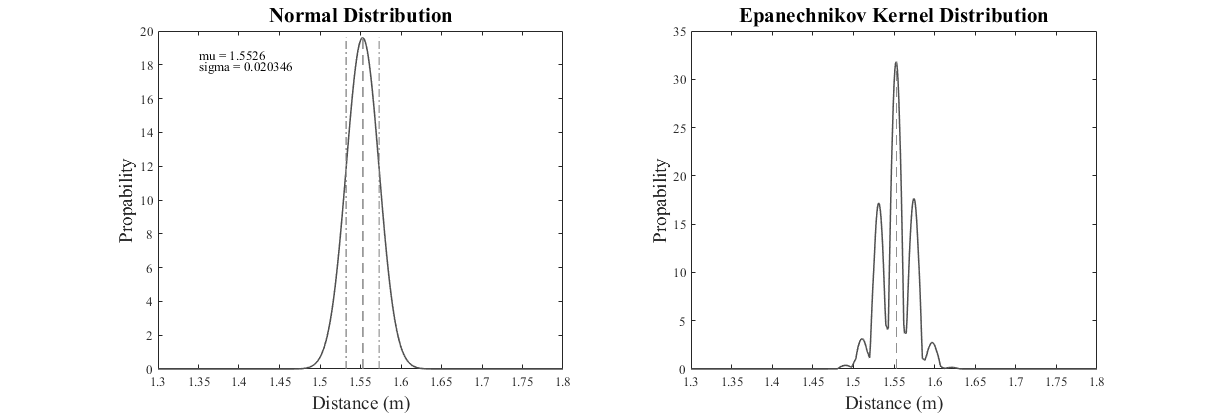}
    \vspace*{10pt}
  \doublecaption[Normal and Epanechnikov distributions.]{Normal and Epanechnikov distributions for a 30\,kHz chirp with a SNR of 6\,dB.}
  \label{Fig:Distributions}
  \vspace{10pt}
\end{figure}

Monte Carlo simulations are performed to test the autocorrelation efficiency when noise is added to a chirp with different signal bandwidths. In a room with an absorption coefficient of 0.9, a single sensor receiver is positioned at a distance of 1.553\,m from the source. The Monte Carlo simulation process is depicted in Fig. \ref{Fig:MonteCarloOverview}. The first two steps are conventional to the source image method: the room impulse response at the microphone is calculated and convolved with the transmitted audio signal. In step 3 white noise is added to create  sound signals with different SNR levels at the receiver. The next step consists of applying a 1\,ms window to the noisy signal, mimicking the wake up time of the mobile nodes. The final step of the simulation involves  autocorrelating the emitted sound signal with the calculated, received sound signal, selecting the index of the correlation maximum and calculating the corresponding distance.  The white noise addition, windowing and correlation processes are repeated for \mbox{10 000} times. Post-processing consists of fitting the acquired distances to a normal distribution and an Epanechnikov Kernel distribution. The latter is chosen due to its optimal performances in a mean square error sense \cite{Epanechnikov}. It shows a better smoothed kernel density estimate in the case of non-Gaussian distributions. Fig. \ref{Fig:Distributions} illustrates the Gaussian and Epanechnikov distributions in the case of a 30\,kHz bandwidth and an SNR of 6\,dB. In the latter, we find secondary peaks offset by a single or multiple wavelengths from the calculated distance peak. In general, the Gaussian distribution is good measurement for the precision and
accuracy.

Table \ref{table:MonteCarlo} shows the accuracy ($\varepsilon$) and precision ($\sigma$) calculated from the Gaussian distribution of five different chirps at increasing signal to noise ratios. The lower frequency is fixed to 25\,kHz, the upper frequency depends on the chosen chirp bandwidth, which is increased in steps of 10\,kHz. As the attenuation of sound propagation is quasi linear with the frequency \cite{Vorlander2008}, we chose descending chirps, i.e. the further away from the sound source, the lower (and less attenuated) the signal within the wake-up window will be. 

Two conclusions can be drawn from Table \ref{table:MonteCarlo}. The bandwidth dependency performs as expected, as increasing the bandwidth improves both accuracy and precision. The influence of additive white noise on the accuracy is minimal. Moreover, the lower chirp bandwidths are less sensitive to the decreasing SNR's, both on accuracy and precision. For noisy environments, better performance is achieved by using these lower chirp bandwidth signals.

\begin{table}[!htb]
\centering
\vspace*{5pt}
\doublecaption{Standard deviation and difference between actual distance and mean value ($\varepsilon$) from Monte Carlo Simulations with a single microphone at a distance of 1.553 m.}
\label{table:MonteCarlo}
\vspace*{5pt}
\renewcommand{\arraystretch}{1.5}
\begin{adjustbox}{width=0.55\textwidth}
\centering
\begin{tabular}{llcccccc}
\toprule
 \textbf{SNR (dB)}&  & \textbf{20} & \textbf{10} & \textbf{6} & \textbf{3} & \textbf{1} & \textbf{0} \\ \hline
35\,khz - 25\,khz & $\varepsilon$ & 0.044 & 0.061 & 0.063 & 0.061 & 0.049 & 0.007 \\ \cline{2-8} 
10 kHz BW & $\sigma$ & 0.024 & 0.043 & 0.058 & 0.092 & 0.277 & 0.573 \\ \hline
45\,khz - 25\,khz & $\varepsilon$ & 0.001 & 0.011 & 0.008 & 0.008 & 0.029 & 0.068 \\ \cline{2-8} 
20 kHz BW & $\sigma$ & 0.004 & 0.021 & 0.31 & 0.042 & 0.300 & 0.565 \\ \hline
55\,khz - 25\,khz & $\varepsilon$ & 0 & 0 & 0 & 0.001 & 0.020 & 0.080 \\ \cline{2-8} 
30 kHz BW & $\sigma$ & 0.001 & 0.014 & 0.020 & 0.042 & 0.316 & 0.652 \\ \hline
65\,khz - 25\,khz & $\varepsilon$ & 0.001 & 0.002 & 0.004 & 0.005 & 0.029 & 0.109 \\ \cline{2-8} 
40 kHz BW & $\sigma$ & 0.001 & 0.006 & 0.014 & 0.023 & 0.435 & 0.833 \\ \hline
75\,khz - 25\,khz & $\varepsilon$ & 0.001 & 0.001 & 0.001 & 0.001 & 0.053  & 0.138 \\ \cline{2-8} 
50 kHz BW & $\sigma$ & 0.001 & 0.004 & 0.008 & 0.099 & 0.550 & 0.887 \\ 
\bottomrule
\end{tabular}%
\end{adjustbox}
\end{table}
Measurements performed in a non-anechoic chamber show that the maximum frequency is limited to 45\,kHz, although the datasheet of the SPU1410LR5H states that the microphone response can go as high as 75\,kHz. The simulation results from above show that with a limited chirp bandwidth of 20\,kHz (45\,kHz downto 25\,kHz), the accuracy and precision remain adequate to perform distance calculations.

\subsection{Room Characteristics Simulations}
\begin{figure}[!htb]
   \centering
   \subfloat[][]{\includegraphics[width=.48\textwidth]{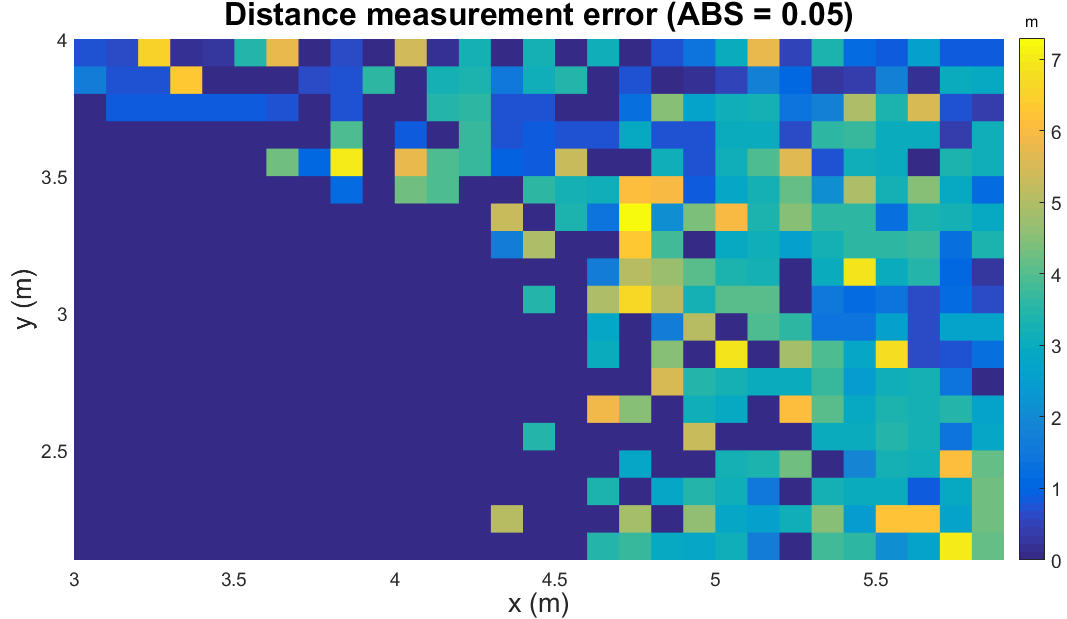}\label{Fig:ABS005} }\quad
   \subfloat[][]{\includegraphics[width=.48\textwidth]{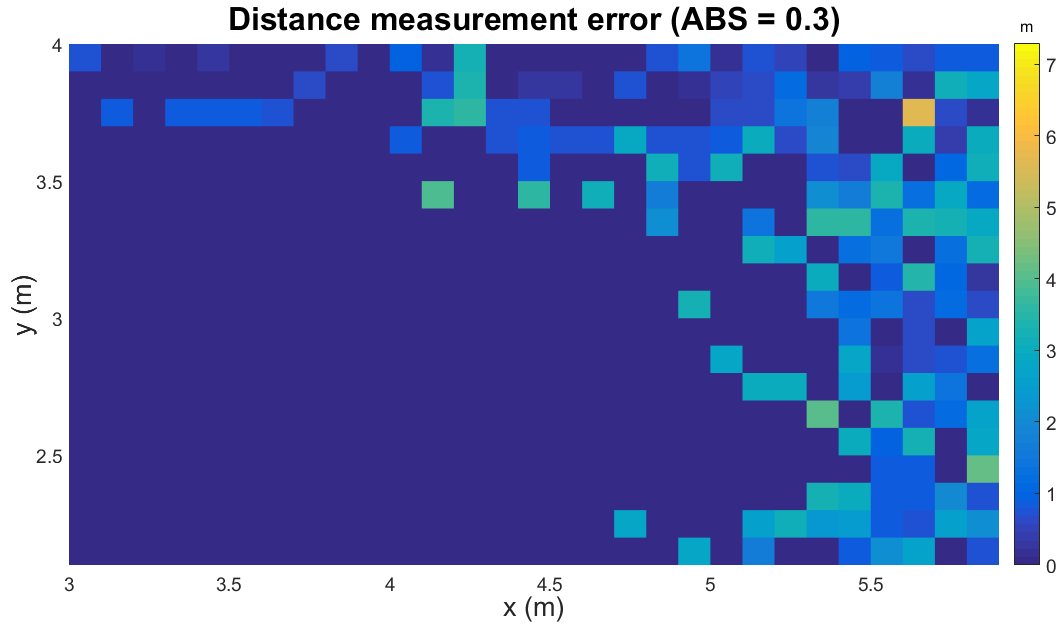}\label{Fig:ABS03}}\\
   \subfloat[][]{\includegraphics[width=.48\textwidth]{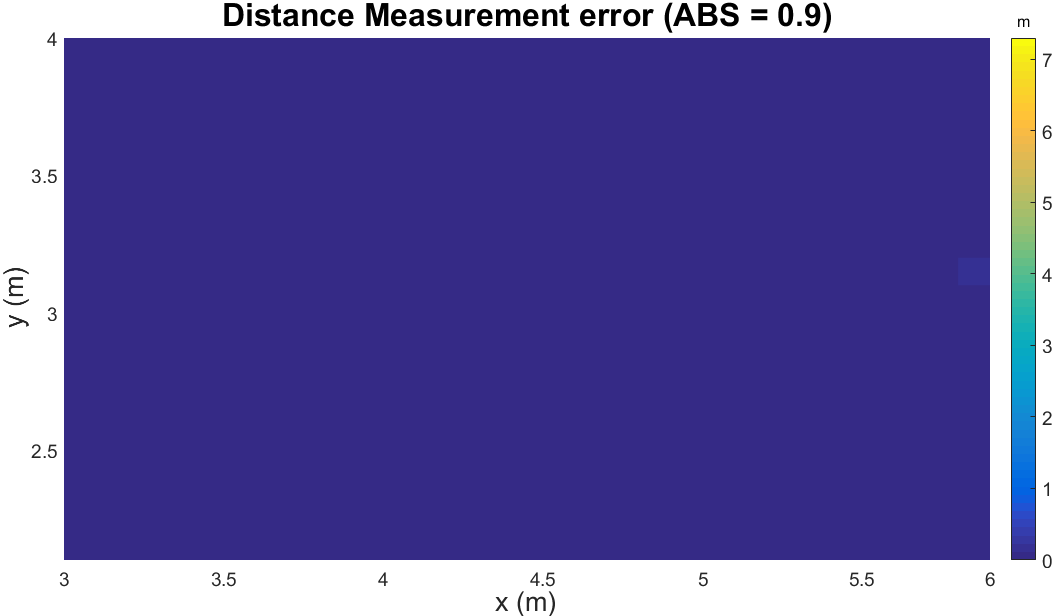}\label{Fig:ABS09}}\quad
   \subfloat[][]{\includegraphics[width=.48\textwidth]{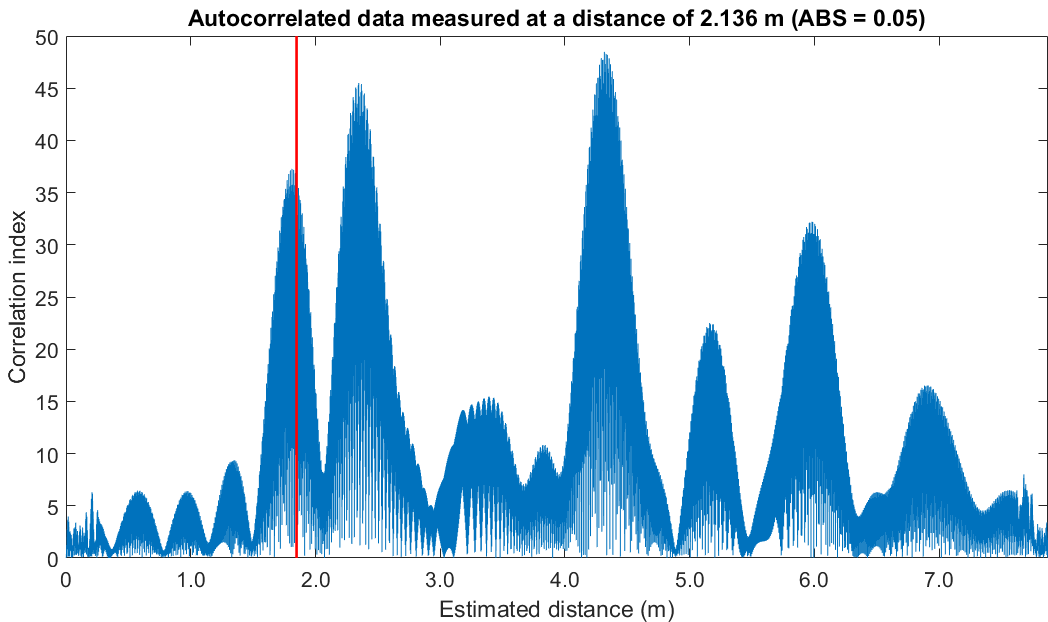}\label{Fig:Crosscorrelation}}
   \vspace*{10pt}
   \doublecaption[Heatmaps on the distance simulation error.]{Heatmaps on the distance simulation error of the room with absorption coefficients (a) $\alpha = 0.05$, (b)  $\alpha = 0.3$ and (c) $\alpha = 0.9 $ . The absolute value of the autocorrelation in case of an ambiguous distance measurement (d) shows the difference between the correct peak (red line) and the maximum.}
   \label{fig:ABS005Unfiltered}
   \setlength{\belowcaptionskip}{-5pt}
\end{figure}

We investigated the impact of the room characteristics, (reflections, diffraction, attenuation, etc.) in the second type of simulations. Three shoeboxes are created, all with a different absorption coefficient: $ \alpha = 0.05$, $ \alpha = 0.3$, and $ \alpha = 0.9$. These absorption coefficients are distinctively chosen and represent respectively an empty room with walls of standard brickwork, fiberboard and acoustic plaster panels \cite{Akustik}. In these rooms, 600 microphones (20x30) are equally spread in one quadrant of the room, with a fixed distance of 10 cm between two sensors or a sensor and a wall. The microphones and the sound source are positioned in the same z-plane, at a height of 1\,m. The P50, P95 and mean distance errors of the pulse compression technique with a chirp bandwidth of 20\,kHz in these three rooms when no noise is added is shown in Table \ref{table:PValues}. The P50 value shows that half of the simulated distances have an error smaller than 3\,cm in the  most real world representative room ($ \alpha_2 = 0.3$). The large difference between the mean and P50 value show that there are a lot of outliers. This is confirmed by the higher P95 values. To visualize what the cause is of these larger errors, a heatmap of the absolute value of the difference between the measured and the actual distance is generated (Fig. \ref{fig:ABS005Unfiltered}). 
\begin{table}[!htb]
\caption{P50, P95 and mean values of the simulated distance calculations in the three rooms when no noise is added to the signal.}
\vspace*{5pt}
\label{table:PValues}
\renewcommand{\arraystretch}{1.5}
\centering
\begin{tabular}{lccc} 

\toprule
\textbf{Absorption coefficient} & \textbf{0.05} & \textbf{0.3} & \textbf{0.9} \\ \hline
Mean                                                & 1.5330        & 0.5462       & 0.0108       \\
P50                                                 & 0.4158        & 0,0293       & 0,0007       \\
P95                                                 & 5.1463        & 3.1489       & 0.0502       \\
\bottomrule
\end{tabular}
\end{table}

It can be seen that there is a considerable, negative effect of the reflections on the accuracy of the proposed system, as expected. The radius in which the ranging performs well decreases as the walls become more reflective. Inspection of the generated correlation data of an erroneous distance simulation, i.e. microphone 522 at a distance of 2.136\,m in a room where $\alpha = 0.05$  (Fig. \ref{Fig:Crosscorrelation}), provides additional insights. The calculated distance corresponding with maximum correlation value is larger than the actual distance, indicated by the red line. Constructive interference of higher frequencies still present in the room can cause correlation peaks larger than the peak generated by the lower, effective distance frequency chirp.

\section{Enhanced Accuracy Solutions}
The previous section showed that utilizing the maximum correlation index as a selection criteria for the distance measurements results in an adequate solution, yet in many situations does not yield the best distance estimate. In most cases, the correct maximum is the first one of a series of local maxima, as the lower frequencies are not present yet due to the descending chirp signal. Three methods are proposed to select this first local peak and enhance the system's accuracy: window functions, peak prominence and delta peak method. The sole constraint of these methods is to keep the processing power as low as possible, as the energy consumption is proportional to it.

\subsection{Method 1: Window Functions}
\begin{figure}
\centering
\begin{minipage}[t]{.48\textwidth}
\centering
\vspace{0pt}
  \includegraphics[width=1.0\textwidth]{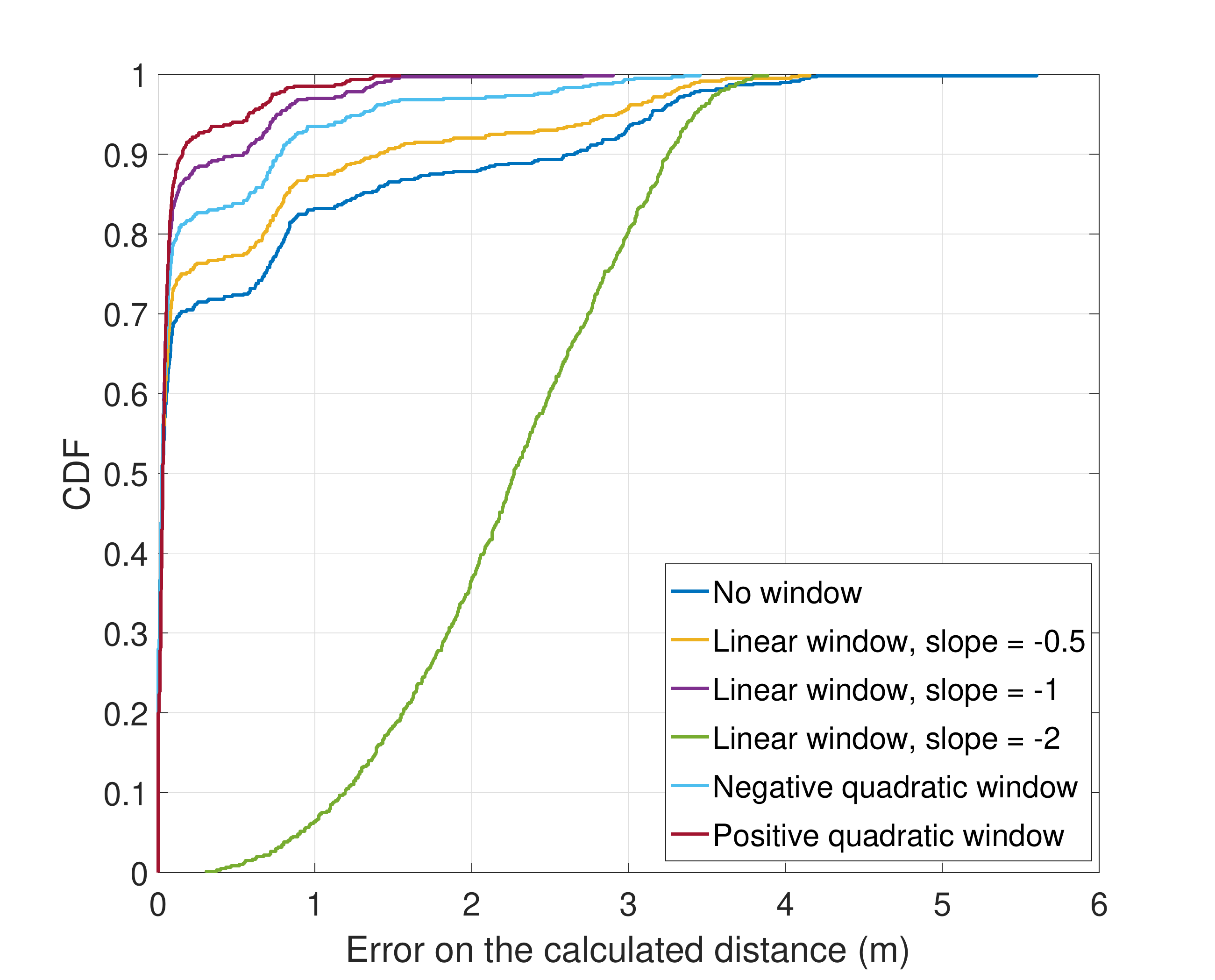}
  \vspace*{5pt}
  \doublecaption[Cumulative density function: linear and quadratic.]{Cumulative density function of the ranging error when linear window functions with different slopes and positive and negative quadratic windows are applied to the correlation data.}
  \label{Fig:CDFplotLinQuad}
\end{minipage}\hfill
\begin{minipage}[t]{.48\textwidth}
\centering
\vspace{0pt}
\includegraphics[width=1.0\textwidth]{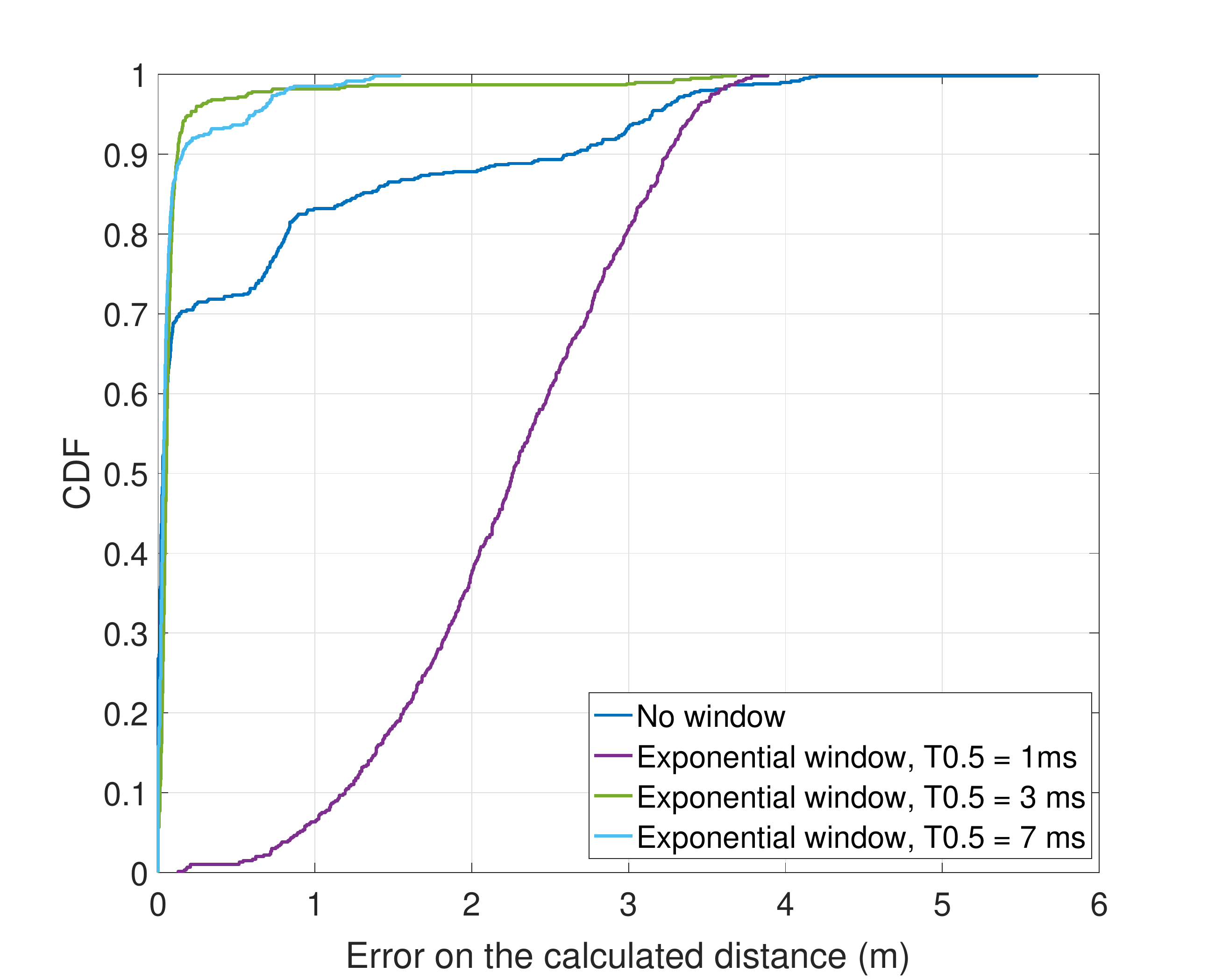}
\vspace*{5pt}
  \doublecaption[Cumulative density function: exponential.]{Cumulative density function of the ranging error when exponential window functions with different bases are applied to the correlation data.}
  \label{Fig:CDFLinQuadExp}
\end{minipage}
\end{figure}

In the first method, window functions are applied to the correlation results, in such a way that the first peaks of the local maxima are increased relatively to the others.  The most straightforward window function is a linearly decreasing function. Fig. \ref{Fig:CDFplotLinQuad} shows the cumulative density function (CDF) plots of the distance error in case of the original, maximum method and when four different linear window functions are applied to the correlation data in the shoe box simulation environment with an absorption coefficient of ($ \alpha = 0.3$). The absolute value of the negative slope of this function can not be larger than 1, as some correlation data will become negative and limit the maximum reachable distance. In general, applying a window function to the correlation results in an improved accuracy of the system. The lowest P50, P95 and maximum distance error (P100) values are obtained with a slope of~-1. We also investigated the potential improvement by applying a quadratic window function of the type $y = \pm ax^2 + bx +c$ to check whether it is better to give more or less weight to the early peaks. Out of the CDF plots in Fig. \ref{Fig:CDFplotLinQuad}, it can be deduced that the positive quadratic function has an even better effect then the linear window function. \newline A faster initial decline consequently increases the influence of the earlier peaks, improving the accuracy. To test the limits of this fast, initial decline, exponential window functions are added: $y = a^{x-b}$ in Fig. \ref{Fig:CDFLinQuadExp}. A good measure of decay in exponential functions is the half-life time ($T_{0.5}$). Smaller decay times ($T_{0.5} = 1\,ms$) result in similar CDF plots as with a steep slope linear function.  Choosing the half-life too large, results in similar CDF plots as the positive quadratic function. The optimal exponential function is the one with a decay time of $T_{0.5} = 3\, ms$, a tenth of the original broadcasted signal. When comparing the optimal exponential window function to the quadratic function, it is clear that the P95 value of this window performs better but the maximum distance errors are more profound. A choice between a more precise or more accurate system can be made here.

\subsection{Method 2: Peak Prominence}
\begin{figure}
\centering
\begin{minipage}[t]{.48\textwidth}
\centering
\vspace{0pt}
\includegraphics[width=1.0\textwidth]{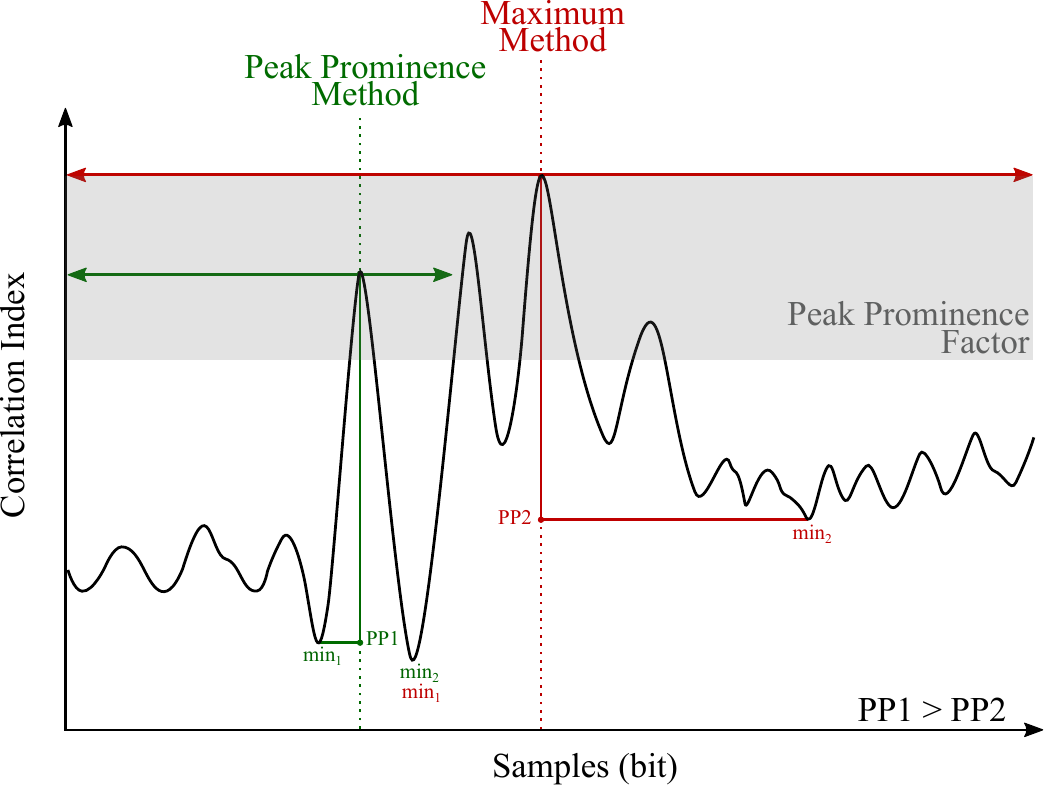}
\doublecaption[PPF and Maximum method comparison.]{Peak Prominence Method compared to the Maximum Method. The correct distance peak is the first of a series of local maxima and correctly appointed to by the peak prominence method.}
\label{Fig:PeakPromMethod}
\end{minipage}\hfill
\begin{minipage}[t]{.48\textwidth}
\centering
\vspace{0pt}
\includegraphics[width=1.0\textwidth]{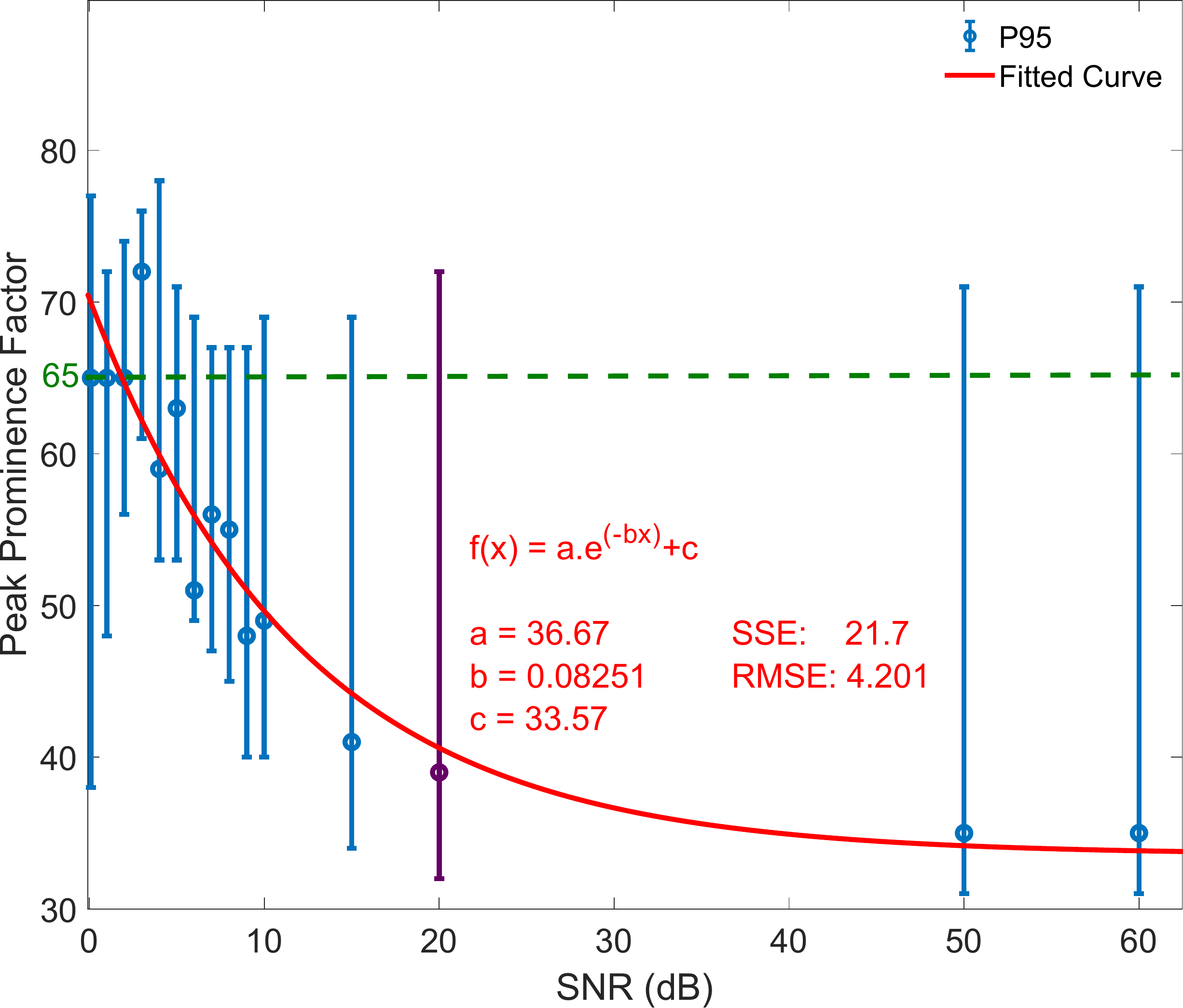}
\doublecaption[Optimal peak performance factor.]{Optimal PPF based on the P95 values for simulations with different white noise SNRs. The cut-off for the distance errors is set to 25\,cm (error bars). An exponential curve can be fit for optimal PPF selection.}
\label{Fig:FittedCurve}
\end{minipage}
\end{figure}

The major discrepancy of the window function method occurs due to the significance it gives to the earlier correlation data. It works well for microphones close to the speaker but as the  difference between the maxima decreases with larger distances, peaks earlier than the correct maximum are chosen, resulting in even larger errors. We further improved the accuracy by searching the local maxima without modifying the pulse compression data.
The prominence of a peak indicates how much the peak stands out due to its intrinsic height and its location relative to other peaks \cite{Chaudry}. It can be calculated as follows: extend a horizontal line from a chosen peak to the left and the right until the line crosses a signal because either there is a higher peak, or it reaches the left or right end of the signal. Find the minimum of the left and right interval (min$_1$ and min$_2$ in Fig. \ref{Fig:PeakPromMethod}). This point can be a valley or a signal endpoint. Calculate the prominence by taking the difference between the height of the peak and the higher minimum of the two intervals. A low, isolated peak can be more prominent than a higher member of a tall range (Fig. \ref{Fig:PeakPromMethod}). This technique is commonly used in topography, in which it represents the elevation of a mountain summit relative to the surrounding terrain, and serves as a criterion to define a separate peak \cite{Press1982} \cite{Summerfield1991}.
The index used for distance calculations is selected by calculating the prominence of all correlation peaks, setting a prominence threshold, the peak prominence factor (PPF), and using the index of the first peak in the array of the prominences larger than the threshold. Both reflections and noise affect the prominence of signals. The influence of the reflections on the accuracy is minimal, as, in a line-of-sight scenario, the correlated maxima of these reflections are positioned later then the original sound signal. Noise on the other hand can reduce the prominence of the correlated peaks, lowering the distinctness of the local maxima, and complicate the PPF determination. 

\begin{figure}[!htb]
 \vspace*{10pt} 
  \centering
    \includegraphics[width=0.85\textwidth]{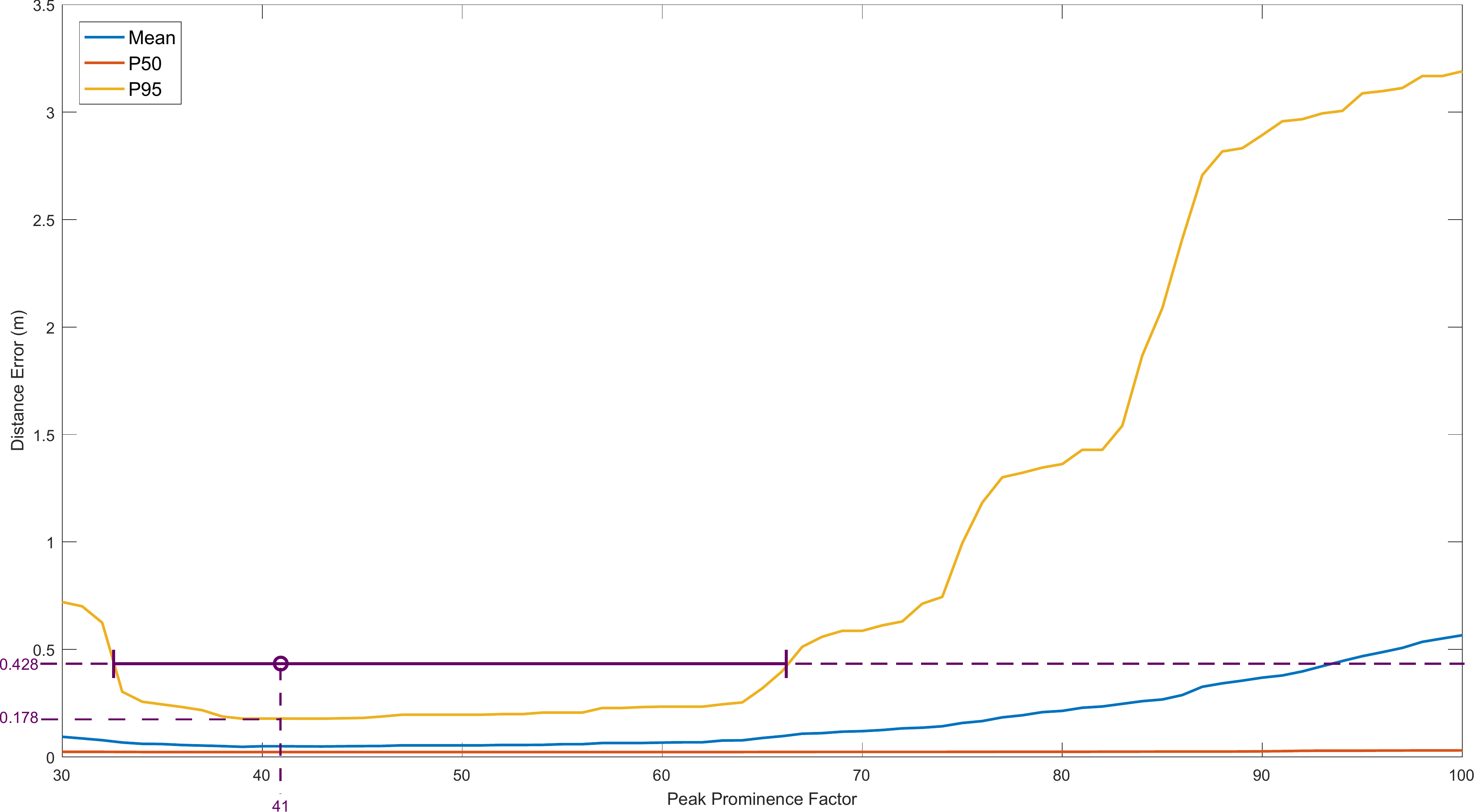}
    \vspace*{10pt} 
  \doublecaption[Mean, P50 and P95 values of the distance error for different PPF.]{Mean, P50 and P95 values of the distance error for different PPF values in a simulation where the white noise SNR is set to 20 dB.}
  \label{Fig:ABS03SNR20}
\end{figure}

In the simulation environment, a clear exponential relationship is found between the SNR and the optimal PPF (Fig. \ref{Fig:FittedCurve}). Determining the SNR requires additional measurements or noise power estimation techniques, both requiring extra energy on processing and power level. Setting a fixed prominence factor can resolve this problem. However, choosing the PPF too low includes erroneous noise peaks to the array, choosing it too high excludes the real distance peaks, resulting in a similar effect as the maximum method. Fig. \ref{Fig:ABS03SNR20} depicts the mean, P50 and P95 values of the distance error at different PPFs in the case of a 20 dB signal to noise ratio. In case of the P95 values, a passband of adequate PPF's can be derived (purple line, here with an upper and lower cut-off value 0.25,/m above the minimum P95 value). These error bandwidths are plotted as bars in Fig. \ref{Fig:FittedCurve} and are proportional to the SNR. A single value (PPF = 65) can be derived from this figure in which the peak prominence method operates adequate over the different SNR values. 
 
\subsection{Method 3: Delta Peak}

\begin{figure}[!htb]
  \centering
    \includegraphics[width=0.85\textwidth]{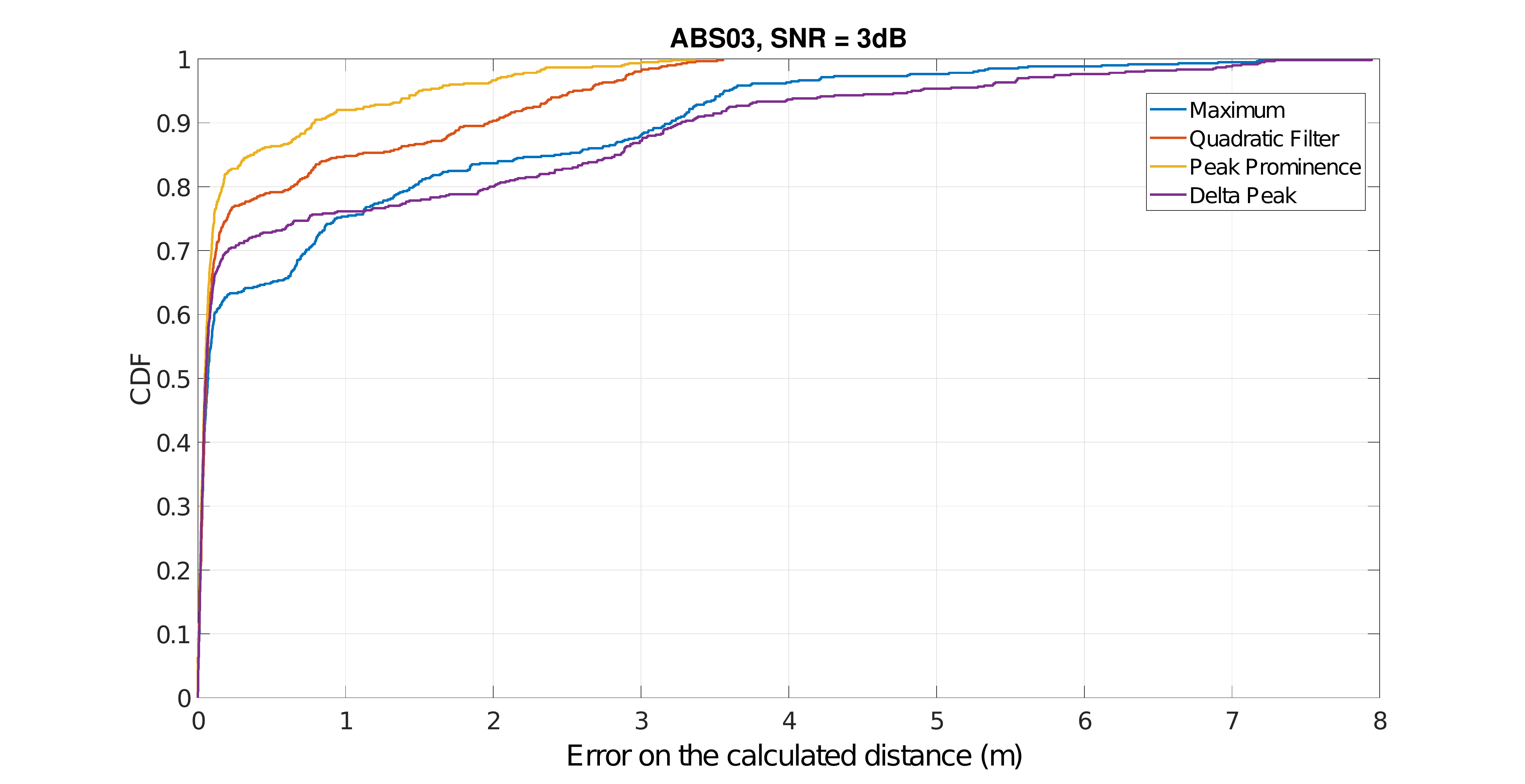}
    \vspace*{10pt} 
  \doublecaption[CDF of proposed optimization methods.]{Cumulative Density Functions of the distance error of the proposed optimization methods in a room with $ \alpha = 0.3$ and  SNR = 3\,dB.}
  \label{Fig:CDFPlotAllMethods}
\end{figure}

\begin{figure}[!htb]
   \centering
   \subfloat[][]{\includegraphics[width=.48\textwidth]{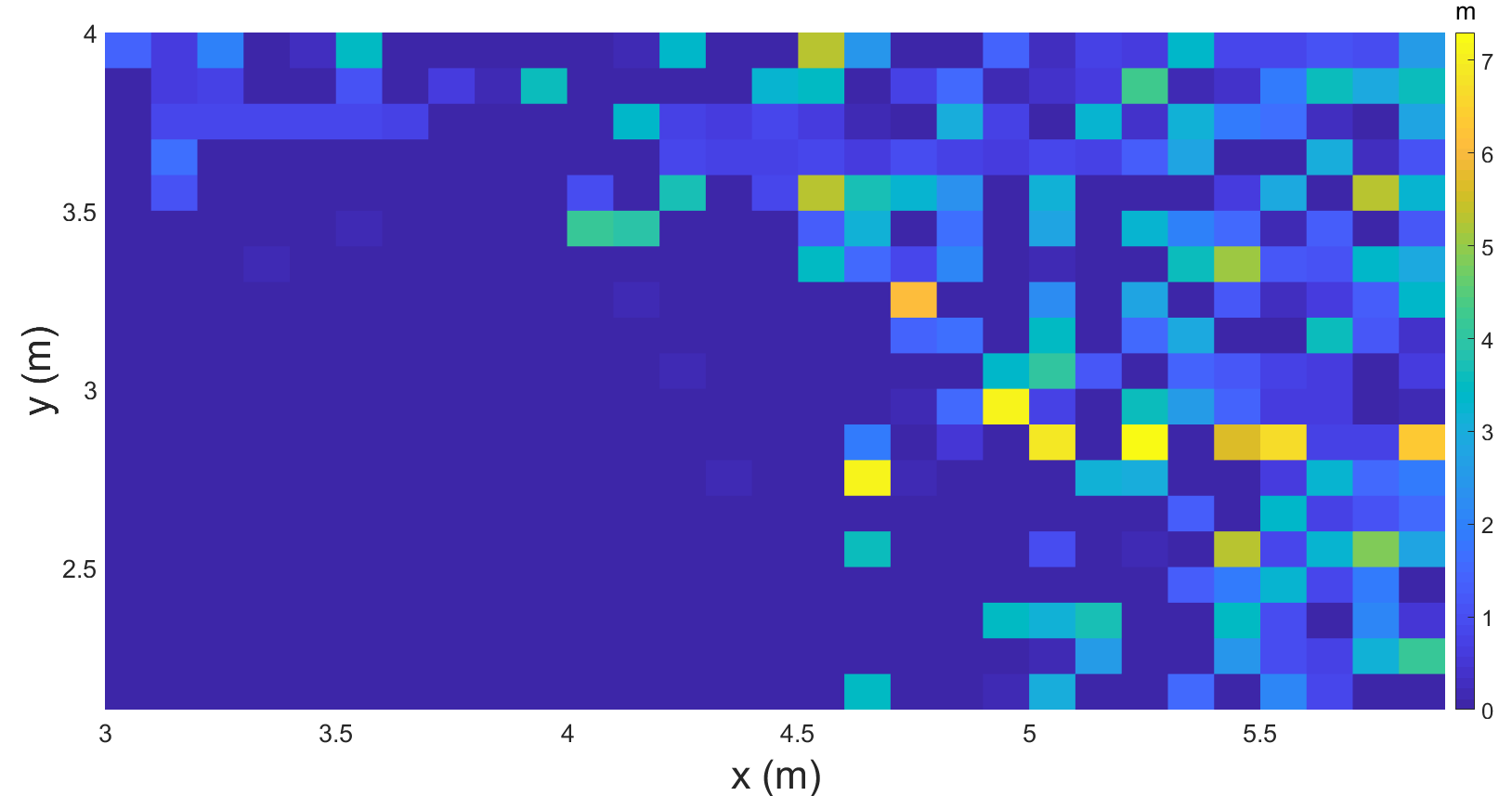}\label{Fig:PColorMaximum} }\quad
   \subfloat[][]{\includegraphics[width=.48\textwidth]{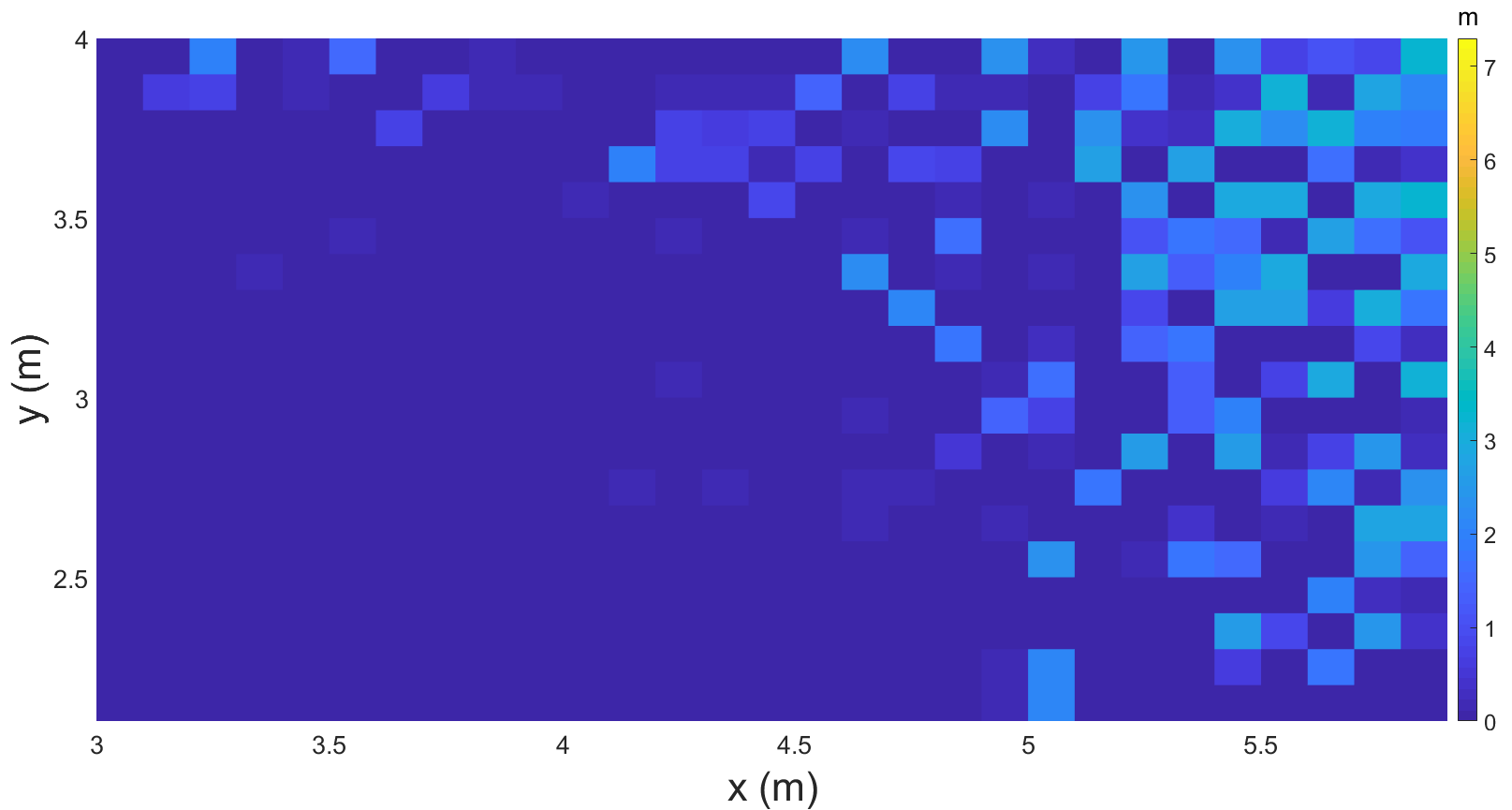}\label{Fig:PColorQuadratic}}\\
   \subfloat[][]{\includegraphics[width=.48\textwidth]{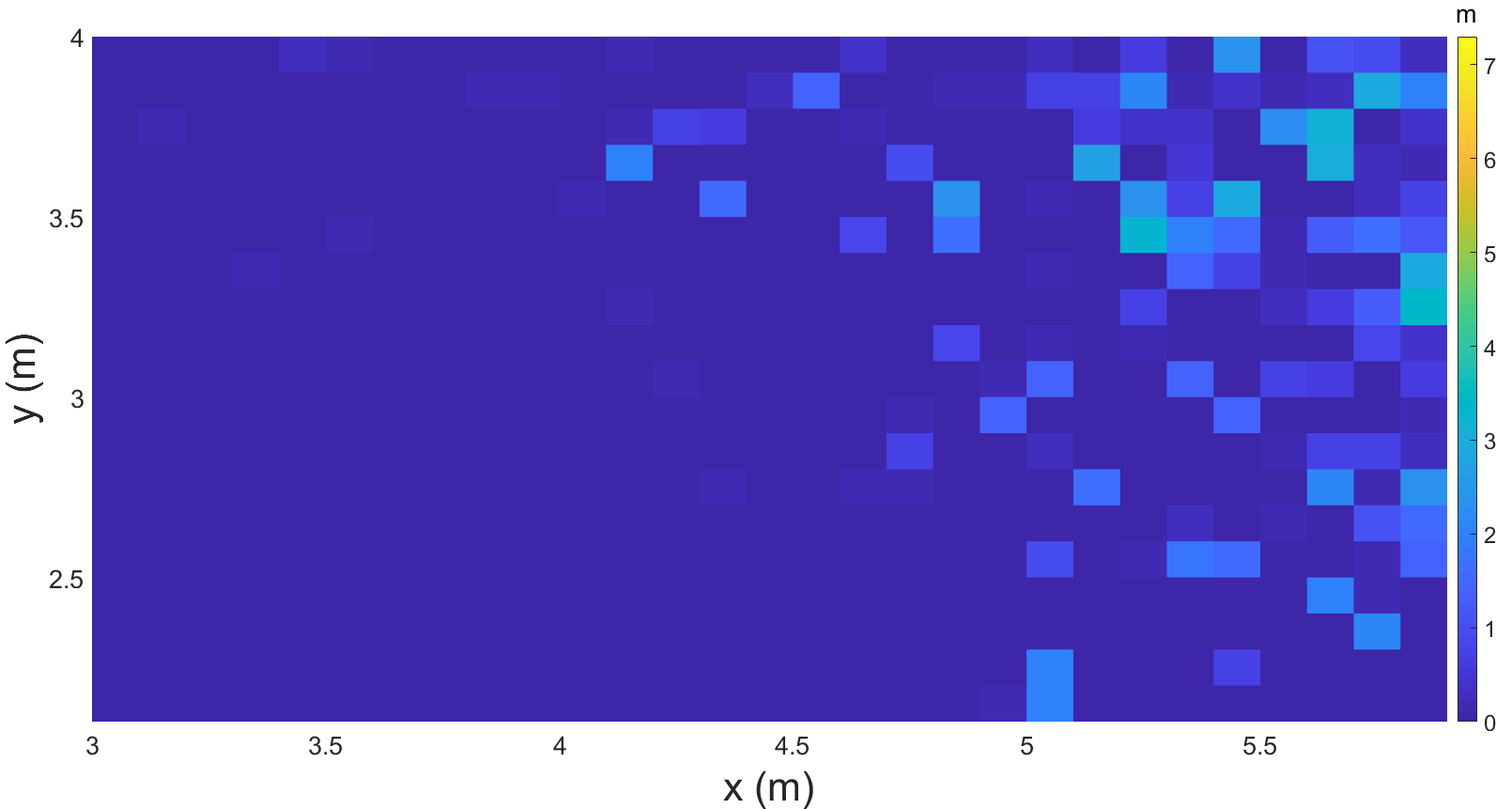}\label{Fig:PColorPeakProm}}\quad
   \subfloat[][]{\includegraphics[width=.48\textwidth]{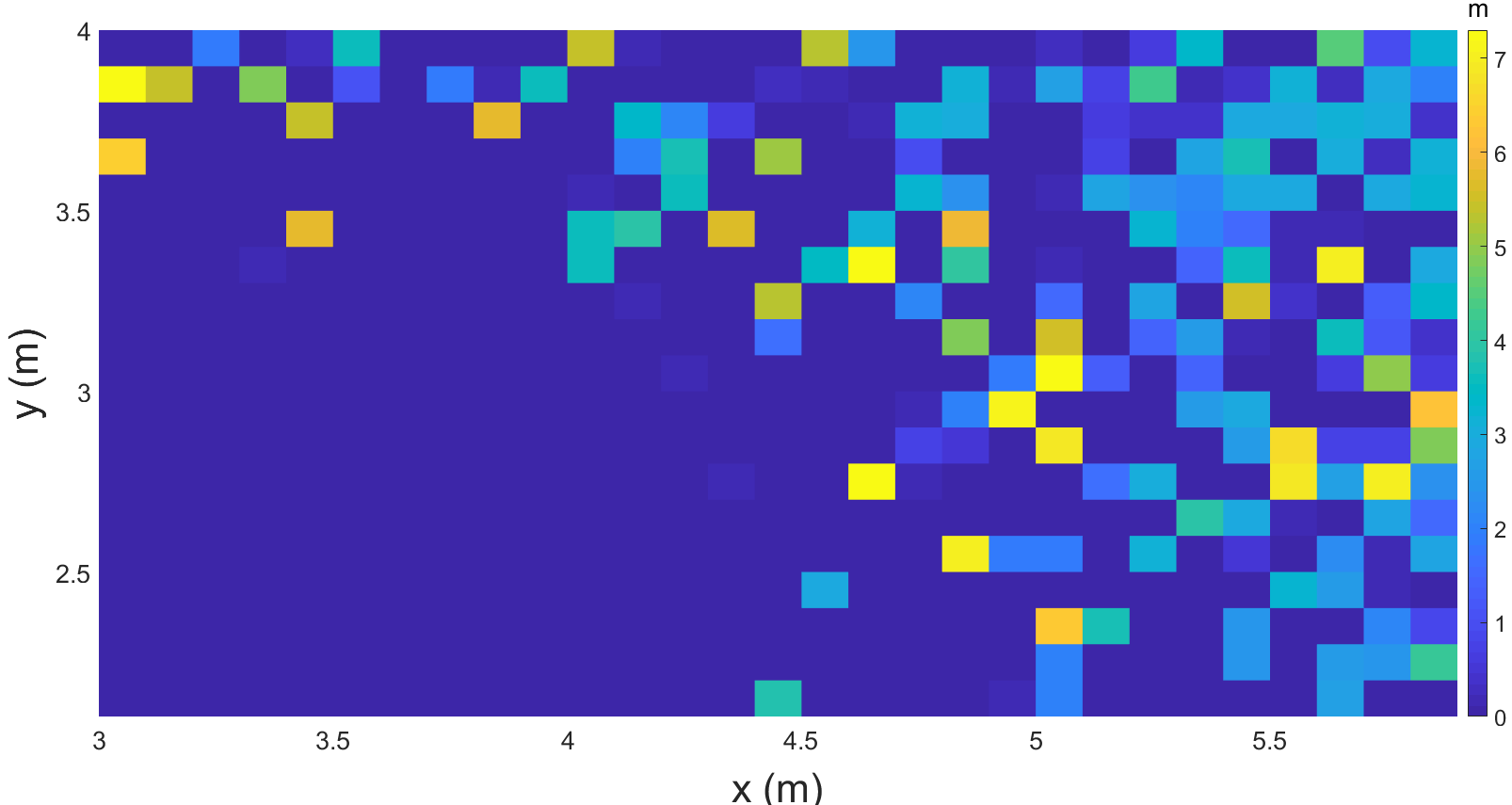}\label{Fig:PColorDeltaPeak}}
   \doublecaption[Heatmaps of the distance error of the different enhanced accuracy solutions.]{Heatmaps of the distance error of the different enhanced accuracy solutions in a simulation environment with absorption coefficient $\alpha = 0.3$ and the SNR = 3\,dB. With (a) the Maximum, (b) the Positive Quadratic Filter, (c) Peak Prominence  and (d) Delta Peak Method.}
   \label{fig:PColorOverview}
   \setlength{\belowcaptionskip}{-10pt}
\end{figure}

We explore the delta peak method as an alternative to improve the system's accuracy. In this approach, the difference between two consecutive local maxima is calculated and the peak following the largest positive difference is selected for further distance estimations. As in the peak prominence method, the correlated data is not altered and the advantage over this latter method is that it is computationally less complex. However, the complexity reduction impairs the accuracy. This can be seen on the CDF of the original and adapted methods for a room with $ \alpha = 0.3$ and a SNR of 3\,dB, charted in Fig. \ref{Fig:CDFPlotAllMethods}. E.g. 63\% of the delta peak distance calculation errors is smaller than 10\,cm, in comparison to 56\% with the maximum method, 68\% with the quadratic window method en 73\% with the peak prominence method. Additionally, the robustness against reflective room characteristics is the lowest of all techniques. Similar to the maximum method, the large delta values due to positive interference imply the wrong index, lowering the system's accuracy. The delta-peak heatmap in Fig. \ref{fig:PColorOverview} shows these outliers close to the corners and walls of the simulation environment.\newline
Table \ref{table:SNRInfluence} represents the mean, P50 and P95 distance error of all proposed methods when different levels of white noise are added. The peak prominence approach has the highest accuracy, even at a very low SNR. Fixing the local peak threshold resolves in a similar, negative SNR-dependency as the other methods. Of the two remaining methods, the positive quadratic window function performs the best.

\begin{table}[!htb]
\vspace*{5pt}
\caption{SNR influence on the distance error (m) of the proposed methods.}
\vspace*{5pt}
\label{table:SNRInfluence}
\renewcommand{\arraystretch}{1.5}
\centering
\begin{adjustbox}{width=0.65\textwidth}
\begin{tabular}{llccccccc}
\toprule
\textbf{METHOD}                      & \textbf{SNR (dB)}              & \textbf{0.1}               & \textbf{1}                 & \textbf{3}                 & \textbf{6}                 & \textbf{10}                & \textbf{20}                & \textbf{60}                \\ \hline
\textbf{Maximum} & \textit{\textbf{Mean (m)}} & 1.223 & 1.149 & 0.879 & 0.742 & 0.626 & 0.572 & 0.547 \\
 & \textit{\textbf{P50 (m)}} & 0.105 & 0.107 & 0.074 & 0.054 & 0.041 & 0.036 & 0.030 \\
 & \textit{\textbf{P95 (m)}} & 5.776 & 5.024 & 4.039 & 3.429 & 3.275 & 3.205 & 3.149 \\
\textbf{Quadratic Window} & \textit{\textbf{Mean (m)}} & 0.673 & 0.590 & 0.424 & 0.223 & 0.103 & 0.093 & 0.090 \\
 & \textit{\textbf{P50 (m)}} & 0.078 & 0.074 & 0.054 & 0.042 & 0.036 & 0.030 & 0.031 \\
 & \textit{\textbf{P95 (m)}} & 2.844 & 2.828 & 2.754 & 1.563 & 0.663 & 0.602 & 0.581 \\
\textbf{Peak Prominence} & \textit{\textbf{Mean (m)}} & 0.465 & 0.375 & 0.262 & 0.160 & 0.099 & 0.094 & 0.086 \\
 & \textit{\textbf{P50 (m)}} & 0.068 & 0.063 & 0.050 & 0.037 & 0.030 & 0.024 & 0.022 \\
 & \textit{\textbf{P95 (m)}} & 2.340 & 2.085 & 1.490 & 0.833 & 0.358 & 0.302 & 0.263 \\
\textbf{Delta Peak} & \textit{\textbf{Mean (m)}} & 1.365 & 1.229 & 0.888 & 0.556 & 0.342 & 0.268 & 0.264 \\
 & \textit{\textbf{P50 (m)}} & 0.121 & 0.098 & 0.055 & 0.041 & 0.030 & 0.025 & 0.024 \\
\textbf{} & \textit{\textbf{P95 (m)}} & 5.509 & 5.003 & 4.499 & 3.382 & 3.057 & 2.678 & 2.746 \\
\bottomrule
\end{tabular}
\end{adjustbox}
\vspace{-10pt}
\end{table}

\newpage
\section{Experimental Validation: Results and Discussion.}\label{sec:results}

We have built a low-power setup to test the proposed system design, the accompanied pulse compression technique and the improved accuracy methods in a real-life environment. The set-up focuses on the key acoustic components of the system described in section \ref{ss:systemOverview}. The aforementioned acoustic transmitter and receiver are realized in hardware and the RF-based wake up is implemented through a cable link between the two entities for proof of concept validation.

\subsection{Experimental Setup and Hardware Prototype.}
\begin{figure}[!ht]
   \centering
   \subfloat[]{\includegraphics[width=.98\textwidth]{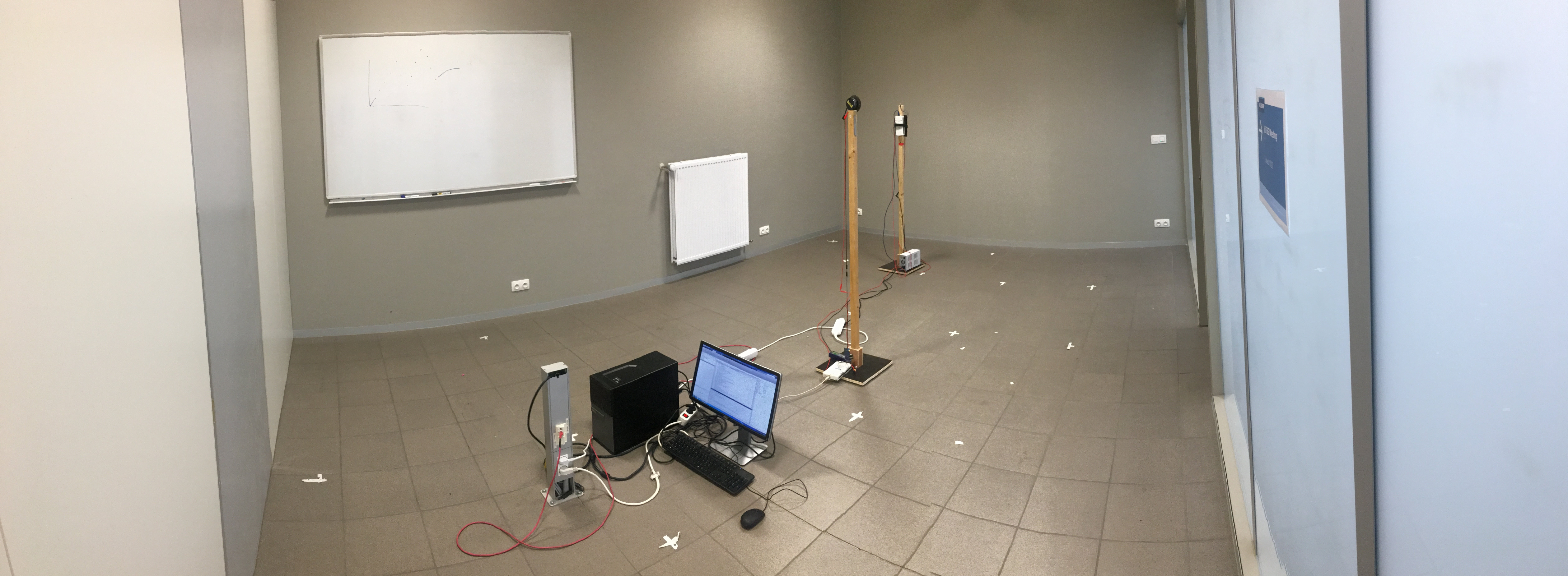}\label{Fig:PhotoSystem} }\\
   \vspace{5pt}
   \subfloat[]{\includegraphics[width=.65\textwidth]{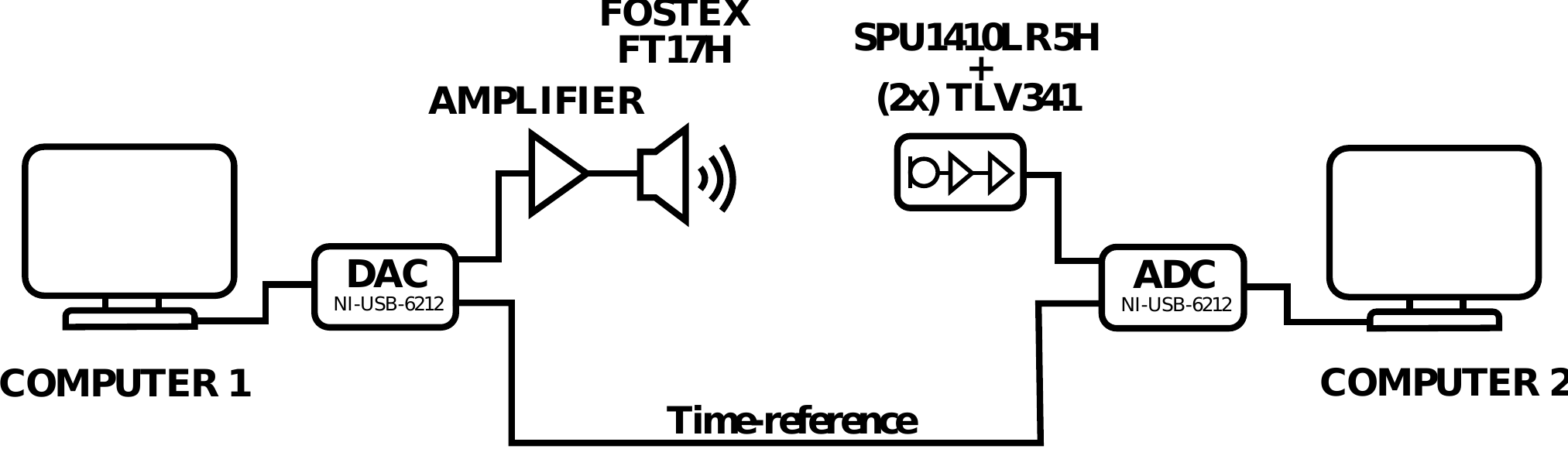}\label{Fig:SimplifiedSystemOverview}}
   \vspace*{10pt} 
   \doublecaption[Measurement environment and setup.]{Picture of the setup in measurement environment and its accompanied, simplified system design.}
   \label{fig:SystemOverviews}
\end{figure}

Fig. \ref{fig:SystemOverviews} depicts the system in an empty 6\,x\,4.27\,x\,3.41\,m room in which the walls consist of plaster-wood and glass, the floor of ceramic tiles and the ceiling of rock wool on solid backing. Three RT60 measurements were performed to test the reverberation time of this room. The average value and uncertainty at the different frequencies can be found in Table \ref{table:RT60Values}. \footnote{The given RT60 values are measured specifically in the audible domain. The RT60 values in the ultrasonic domain will be lower, as the reverberation time is inversely proportional to the frequency.}

To receive the ultrasonic sound signals at low-power, dedicated hardware has been designed based on an ultra low-power acoustic array \cite{Thoen}. Ultrasonic MEMS microphones (SPU1410LR5H \cite{SPU1410}) are used as a sound transducer. The advantages of these MEMS microphones over ultrasonic piezo-elements are their omnidirectional response, small form factor and high bandwidth. The two opamps (TLV341 \cite{TLV341}) have a large gain bandwidth product, to oppose the low amplitude signals coming from the microphone. The acoustic measurements on the MEMS and amplification circuit show a maximum detectable frequency of 45\,khz. Active filters are added to the cascading opamps to narrow the boosted signals to the limited bandwidth in the ultrasonic domain (25kHz to 45\,khz). A fixed LDO voltage regulator is added as a supply for the MEMS microphone and as an input offset voltage to guarantee a rail-to-rail output. The output of this ultrasonic receiver is then sampled and used as an input signal for the pulse compression. In this paper, the acoustic data is sampled with a NI-USB-6212 DAQ \cite{NIUSB6212}. The sample frequency and data resolution are adapted to mimic the ADC used in common microcontrollers, (196 kS/s and 12 bit).
Further processing of the collected data is done in a Matlab environment.\newline
\begin{table}[!htb]
\vspace*{10pt}
\caption{Power measurements of the proposed hardware. Power measurements of the LDO, MEMS and OPAMPS are performed with a 4 points measurement on a Hameg HM8112-3 precision multimeter.}
\label{table:PowerMeas}
\vspace*{10pt}
\renewcommand{\arraystretch}{1.5}
\centering
\begin{adjustbox}{width=0.75\textwidth}
\begin{tabular}{lccccc}
\toprule
 & \multicolumn{3}{c}{\textbf{Acoustic Front-End}} & \textbf{Edge-processing} \\
\cmidrule(lr){2-4}\cmidrule(lr){5-5}
 & \begin{tabular}{@{}c@{}}\textbf{LDO +} \\ \textbf{MEMS}\end{tabular} & \textbf{OPAMP 1} & \textbf{OPAMP 2}   & 
 \begin{tabular}{@{}c@{}}\textbf{ADC (nRF52)} \\ \textbf{datasheet}\end{tabular} & \textbf{\textit{Total}} \\
 \cmidrule(lr){2-6}
Active Current ($\mu$A) & 113.1 & 81.5 & 81.5 & 300 & \textit{576.1} \\
Passive Current ($\mu$A) & 0.03 & 0 & 0 & 1.9 &  \textit{1.93} \\
Active Power (nW) @1\,ms & 407.2 & 293.4 & 293.4 & 1080 & \textit{2074.0} \\
Passive Power (nW) @ 999\,ms & 107.9 & 0 & 0 & 6833.2 & \textit{6940.9} \\
Total Power (nW) & 515.1 & 293.4 & 293.4 & 7913.2 &  \textit{\textbf{9014.9}}\\
\bottomrule
\end{tabular}
\end{adjustbox}
\vspace{-10pt}
\end{table}

The components above are specifically chosen to address the obliged low-power prerequisites. The LDO and amplifiers are equipped with power down pins, reducing the quiescent power consumption when the ranging system is not activated. Table \ref{table:PowerMeas} summarizes the power usage of these components and the estimated power usage of a nRF52832's ADC, as described in the datasheet. Due to the short awake time of the mobile node, the power consumption in passive state has a major contribution to the total power consumption. On a standard single coin cell battery (CR2032: 3\,V, 225\,mAh \cite{CR2032}) this system could operate for more than 8.5 years if the receiver would wake up once every second. This equals the battery shelf life. Note that wake-up times of the LDO, microphone, ADC or opamps can  increase the power consumption significantly.\newline
The transmit side consists of the following elements: a DAC, a commercially available amplifier circuit and an ultrasonic speaker. Here, the NI-USB-6212 DAQ is used as a DAC to generate two signals: the 45\,kHz to 25\,kHz chirp and the 'start-sampling'-signal. The latter signal is sent over the cable to the receiver side and consists of a pulse at $T_A$ imitating the RF-wake-up. This pulse initiates a 1ms sampling time at the receiver DAQ. The DAC chirp signal is amplified with a commercially available amplifier circuit, based on a TDA7492 class-D opamp. In house tests have show that it has an amplification bandwidth over the intended 45\,khz. As an ultrasonic sound speaker, the Fostex FT17H  is chosen. Its ultrasonic capabilities comes at a cost, limiting the directionality on a XY-plane to 30$^{\circ}$  at a 25\,kHz sine wave.

    

\begin{table}[!htb]
\vspace*{5pt}
\captionof{table}{Average RT60 values of the room where the measurements were performed. }
\vspace*{5pt}
\label{table:RT60Values}
\renewcommand{\arraystretch}{1.5}
\centering
\begin{adjustbox}{width=0.55\textwidth}
\begin{tabular}{lcccccc} 
\toprule
\textbf{Frequency (Hz)} & \textbf{250} & \textbf{500} & \textbf{1000} & \textbf{2000} & \textbf{4000} & \textbf{8000}\\ \hline
RT60 (s)                                               & 1.17        & 0.94       & 0.78 & 0.78 & 0.79 & 0.59       \\
Uncertainty (\%)                                       & 7.8         & 6.2        & 4.8  & 3.4  & 2.4  & 1.9      \\ 
\bottomrule
\end{tabular}
\end{adjustbox}
\end{table}

\subsection{Ranging Measurements}
\begin{figure}[!ht]
   \centering
   \subfloat[][]{\includegraphics[width=.48\textwidth]{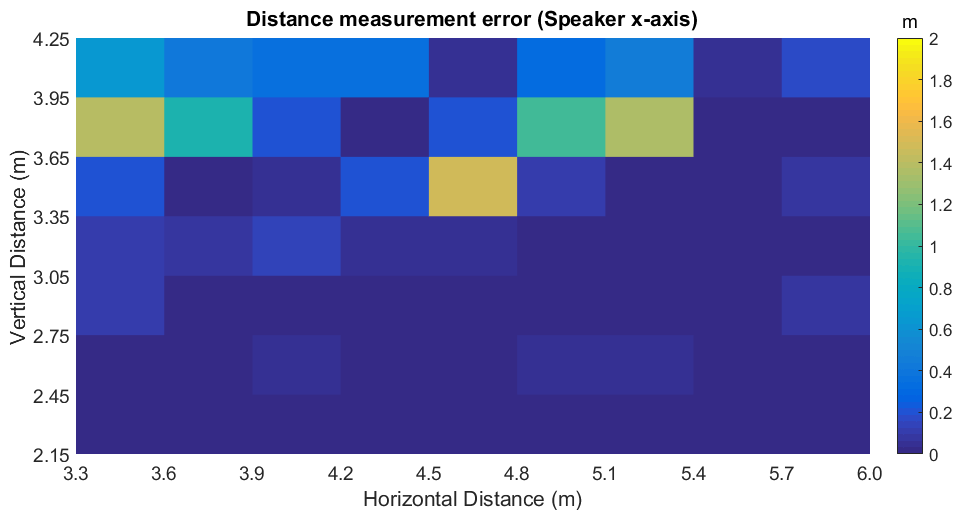}\label{Fig:SpeakerXaxis} }\quad
   \subfloat[][]{\includegraphics[width=.48\textwidth]{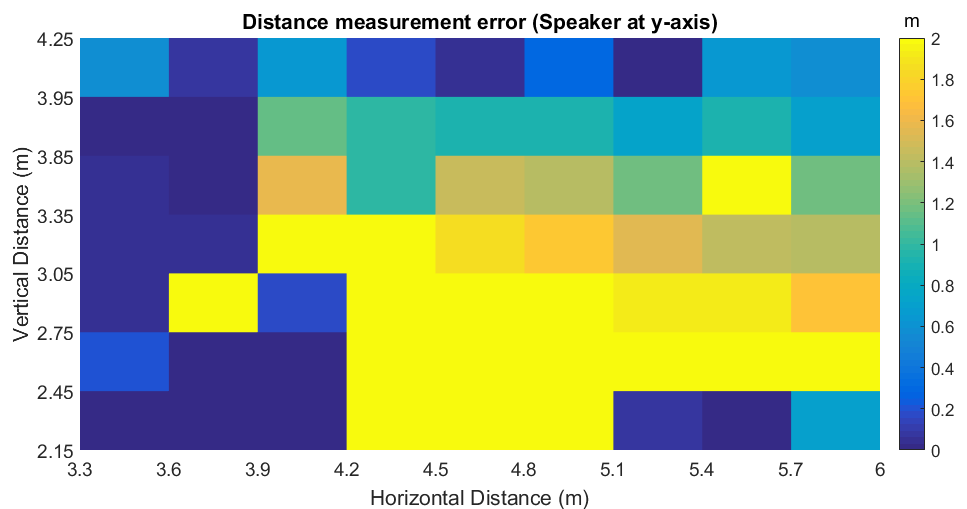}\label{Fig:SpeakerYaxis}}\\
   \subfloat[][]{\includegraphics[width=.48\textwidth]{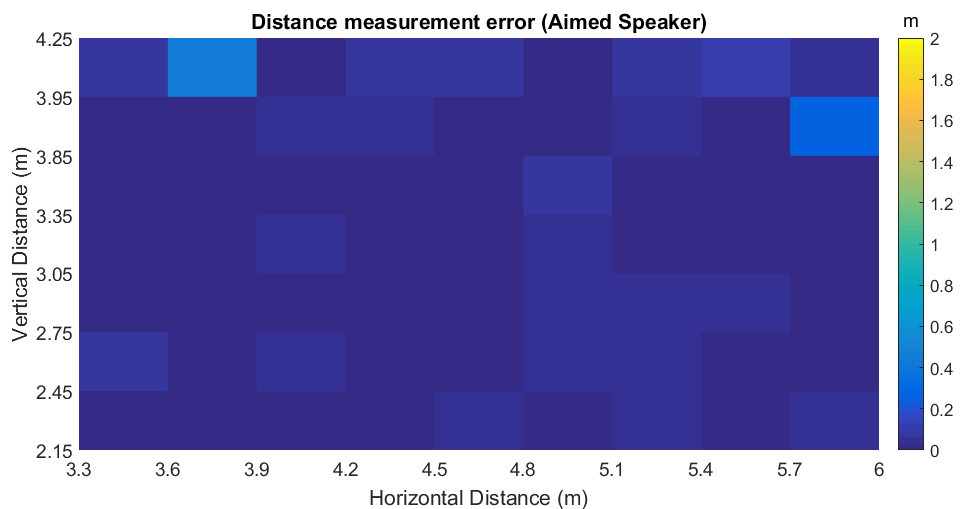}\label{Fig:Aimed}}\quad
   \subfloat[][]{\includegraphics[width=.48\textwidth]{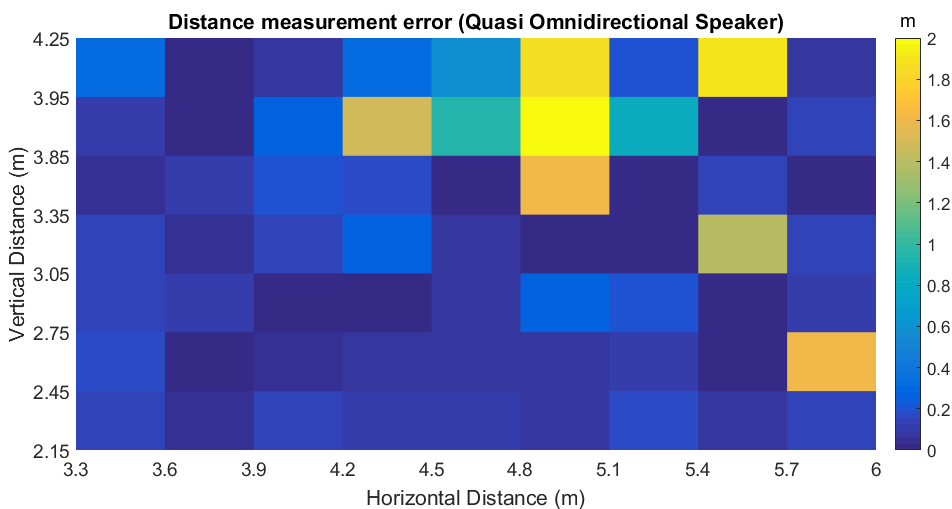}\label{Fig:SpeakerOmnidi}}
   \vspace*{10pt}
   \doublecaption[Heatmap of maximum method in different measurement scenarios.]{Heatmap of the maximum method in different measurement scenarios: when the speaker is aimed towards the x-axis (a), towards the y-axis (b), towards every microphone position individually (c) and towards the z-axis with a semi-sphere (quasi-omnidirectional speaker) (d).}
   \label{fig:sub1}
\end{figure}

\begin{figure}[!ht]
   \centering
   \subfloat[][]{\includegraphics[width=.48\textwidth]{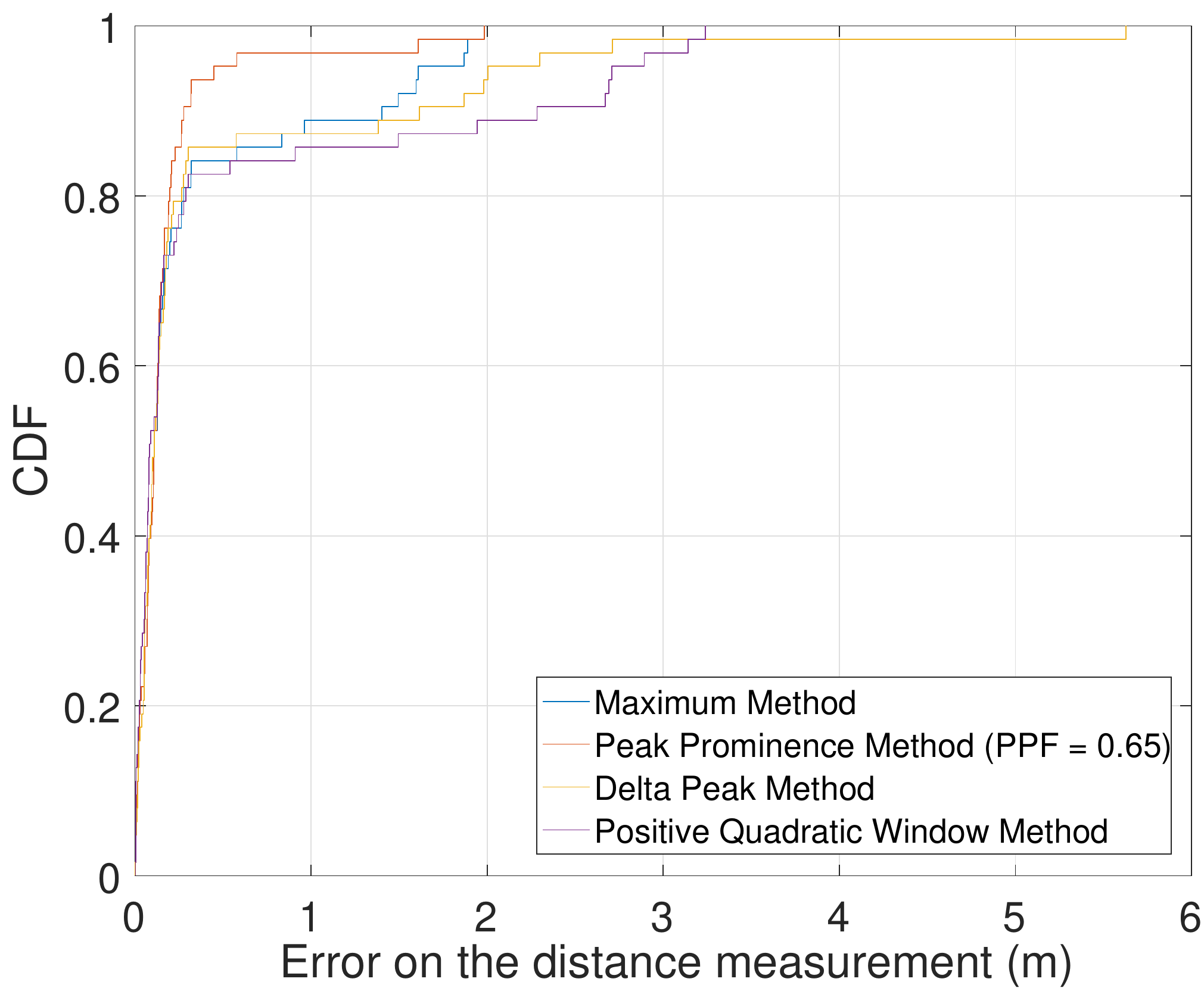}\label{Fig:CDFplotMeasurement5} }
   \subfloat[][]{\includegraphics[width=.48\textwidth]{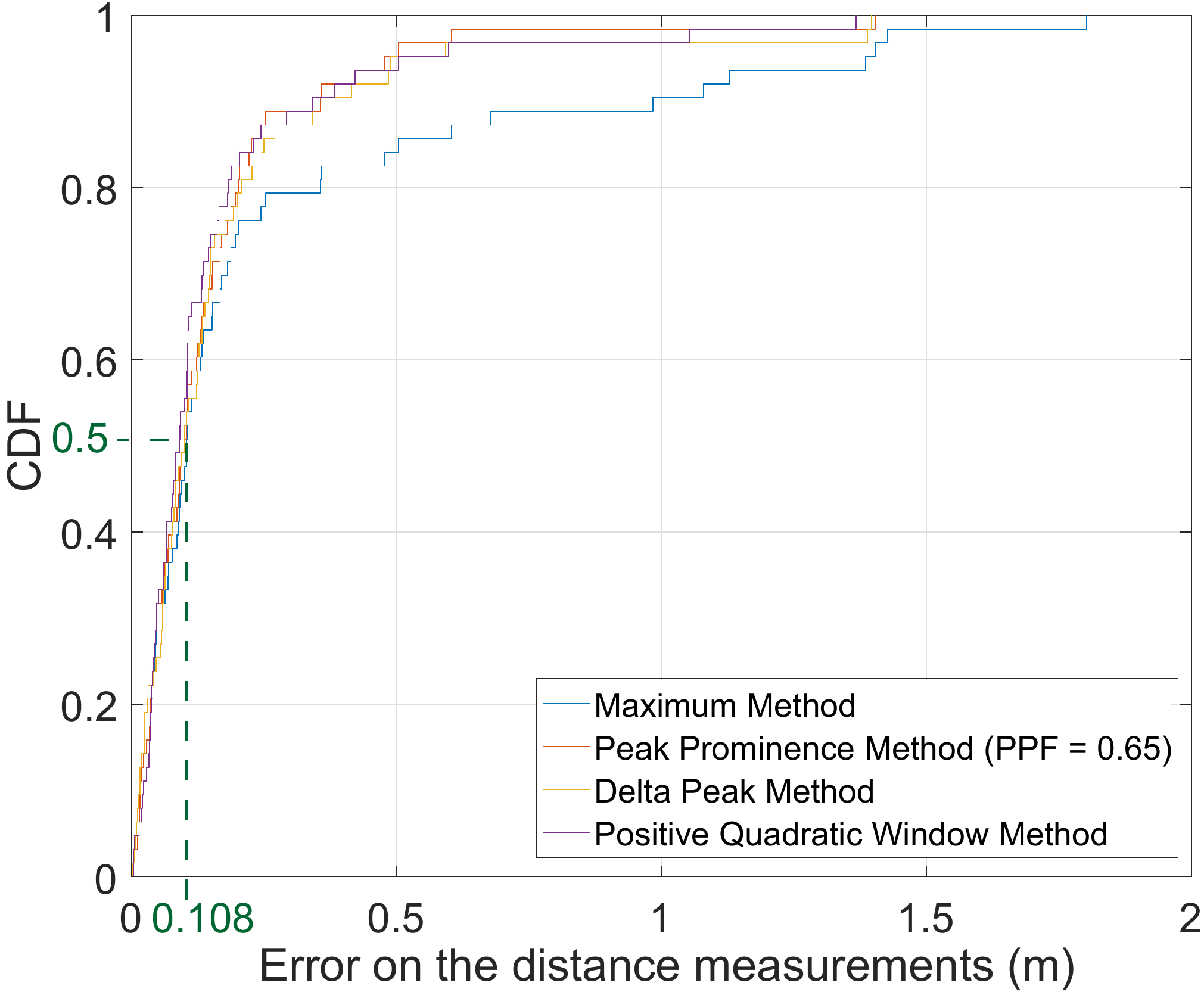}\label{Fig:CDFplotMeasurement6}}
   \vspace*{10pt}
   \doublecaption[CDF plot of the improved accuracy methods.]{CDF plot of the improved accuracy methods on quasi-omnidirectional measurements with (a) a lower and (b) a higher SPL.}
   \label{Fig:CDFplotMeasurements}
\end{figure}

We evaluated the accuracy of the proposed solution by performing acoustic measurements in a quadrant of the room, in which 63 measurement points, with a mutual distance of 30\,cm, were dispersed. Three types of scenarios were tested: with the speaker directed to the x- or y-axis, with the speaker directed to the microphone and with an adapted, quasi-omnidirectional speaker. As in the simulations, the sound source is located at an off-centered position. \newline
The first measurement scenario shows that the signal power received outside the directional speaker's beam is limited, resulting in large accuracy errors, as can be seen in Fig. \ref{Fig:SpeakerXaxis} and \ref{Fig:SpeakerYaxis}. To obtain a quasi-omnidirectional speaker, a semi-sphere is put on a distance from an upwards oriented tweeter, reflecting the sound in all possible directions.  Tests in the audible domain show only a difference of 6\,dB between the maximum and minimum measured sound intensity level due to the structure of the speaker. This method addresses the preeminent scenario in which the directional sound source is directed towards the receiver, and only reflections from the walls in the speaker direction are received by the microphone (Fig. \ref{Fig:Aimed}).\newline
The results of the maximum method applied on these latter,  quasi-omnidirectional speaker measurements are in line with the corresponding simulations (Fig.~\ref{Fig:SpeakerOmnidi}). The accuracy of these measurements is high for microphones close to the sound source. If the distance is increased or the microphone is closer to a wall, the accuracy drops. Fig. \ref{Fig:CDFplotMeasurements} shows that the median of the distance error in this scenario is 0.108\,m. The large P95 value (1.608\,m) reveals again a number of outliers. 

\begin{figure}[!htb]
  \centering
    \includegraphics[width=0.850\textwidth]{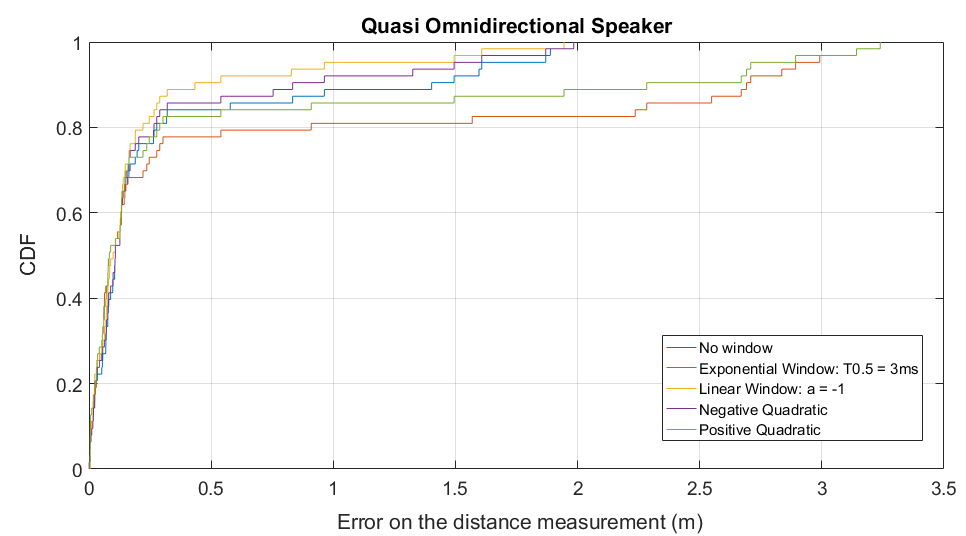}
    \vspace*{10pt}
  \doublecaption[CDF of the proposed window function on measurement data.]{Cumulative Density Function of the proposed window functions on the measurement data performed with the quasi-omnidirectional speaker.}
  \label{Fig:CDFWindowMeasurements}
\end{figure}

We applied the aforementioned improved accuracy methods to two sets of measurement data, representing a low and high SNR scenario. The CDF plot of these two measurements can be found in Figure \ref{Fig:CDFplotMeasurements}. In the higher SPL scenario, all of the proposed methods show the behaviour that is anticipated by the simulations except one: the delta peak method. The smaller amount of outliers in this method result in a lower P100 value than the maximum method and we can conclude that in the higher SPL measurements, all of the proposed methods have a higher accuracy than the original, maximum method. The P50 and P95 values are respectively lower than 10\,cm and 50\,cm.\newline
We noticed a risk when the received sound signal power is reduced. At higher P-values (\textgreater \,P80), the curves do not follow the same path as in the simulation and higher SPL scenarios. In these measurements, the peak prominence and  maximum method have the best P100 (1.984\,m) values. The smaller mean and P95 value of the peak prominence method confirm that this method has a lower amount of larger distance measurements errors making this method more robust than the maximum method.  The delta peak in the measurements method shows a similar behaviour as in the simulations where the large distance errors outliers lead to a method that performs worse than the original maximum method. Comparable results are found for the positive quadratic window method. A more detailed CDF plot of the applied window methods (Fig. \ref{Fig:CDFWindowMeasurements}) shows that the two optimal window functions, the exponential window with a half life of 3\,ms and the positive quadratic function, perform worse than the data without any window. This is opposite to the results from simulations. The linear function performs best. When comparing the correlation data of the window functions on a single position with low accuracy, it is clear that the early correlation peaks in the case of the positive quadratic and exponential windows, are overamplified and larger than the accurate distance peak. This increases the distance error resulting in a poorer distance accuracy.\newline

\section{Conclusions and Future Work}

We have presented and demonstrated a novel hybrid signaling distance measuring system at ultra low power consumption. The system is able to perform cm-accurate distance measurements with sampled and small (196 sampled during 1ms) ultrasonic chirp signals on a restricted energy budget. Monte Carlo and acoustic shoe box simulations with 600 distributed microphones show centimeter accuracy for the synchronized wake-up and pulse compression method close to the sound sources. This accuracy is in line with the current state-of-the-art indoor positioning systems, often performed in artificial environments \cite{Zafari2019}.  We proposed three lightweight and fast methods to improve the accuracy near reflective objects with a limited processing power. The Peak Prominence Method improves the accuracy in the low SNR scenario's with a factor 10 for the P95 values. Experimental verification with an in-house developed ultrasonic receiver validate the enhanced accuracy methods and confirm the low-power acoustic reception and processing, with a power consumption of 2.074\,$\mu$W for a single 1\,ms measurement or over 8.5 years of operation on a CR2032 coin cell battery. This result is in agreement with previous analyses  \cite{Marioli} \cite{Nakahira} demonstrating comparable results with the used autocorrelation method. In our future work we will extend the experimental set-up with the RF signaling. This power consumption is 3 orders of magnitudes more efficient than BLE-indoor positioning technique proposed by \cite{Sadowski2018}.

Next, a calibration solution will be worked out to determine the start-up time of the receiver hardware components, as it may impact both precision and power consumption. \newline
 The presented ranging method can be extended to perform positioning. Extra sound sources should be added to the set-up for positioning purposes in a 2D and 3d environment. This suggest that at the awake time, three signals are received at the same time. Needless to say, identifying which signal comes from which source is the main concern here. A more detailed investigation should reveal which multiple access protocol is suited to address this challenge and existing chirp based techniques \cite{Khyam2} should be tested.
\newpage
\bibliographystyle{unsrt}  
\bibliography{library.bib}

\end{document}